\documentclass[preprint,prd,nofootinbib,tightenlines,amsmath]{revtex4}
\pdfoutput=1
\usepackage{enumerate}
\usepackage{multirow}
\usepackage[utf8]{inputenc}
\usepackage{anysize}
\usepackage{textcomp}
\usepackage{epsfig}
\usepackage{graphics,color} 
\usepackage{verbatim}
\usepackage{afterpage}
\usepackage{amsmath}    
\usepackage{soul}
\usepackage{xcolor,cancel}

\DeclareUnicodeCharacter{00A0}{ }

\usepackage{blindtext}

\catcode`\@=11
\def\sla@#1#2#3#4#5{{%
 \setbox\z@\hbox{$\m@th#4#5$}%
 \setbox\tw@\hbox{$\m@th#4#1$}%
 \dimen4\wd\ifdim\wd\z@<\wd\tw@\tw@\else\z@\fi
 \dimen@\ht\tw@
 \advance\dimen@-\dp\tw@ \advance\dimen@-\ht\z@
 \advance\dimen@\dp\z@
 \divide\dimen@\tw@ \advance\dimen@-#3\ht\tw@
 \advance\dimen@-#3\dp\tw@ \dimen@ii#2\wd\z@
 \raise-\dimen@\hbox to\dimen4{%
 \hss\kern\dimen@ii\box\tw@\kern-\dimen@ii\hss}%
 \llap{\hbox to\dimen4{\hss\box\z@\hss}}}}

\def\cpto{\mathrel {\vcenter {\baselineskip 0pt \kern 0pt
    \hbox{$H_{r.f.}$} \kern 0pt \hbox{$\longrightarrow$} }}}
\def\slashed#1{%
 \expandafter\ifx\csname sla@\string#1\endcsname\relax
{\mathpalette{\sla@/00}{#1}}
\fi}
\def\declareslashed#1#2#3#4#5{%
 \expandafter\def\csname sla@\string#5\endcsname{%
#1{\mathpalette{\sla@{#2}{#3}{#4}}{#5}}}}
 \catcode`\@=12
\declareslashed{}{/}{.08}{0}{D}
 \declareslashed{}{/}{.1}{0}{A}
 \declareslashed{}{/}{0}{-.05}{k}
 \declareslashed{}{/}{.1}{0}{\partial}
 \declareslashed{}{\not}{-.6}{0}{f}

\def\lsim{\mathrel {\vcenter {\baselineskip 0pt \kern 0pt
    \hbox{$<$} \kern 0pt \hbox{$\sim$} }}}
\def\gsim{\mathrel {\vcenter {\baselineskip 0pt \kern 0pt
    \hbox{$>$} \kern 0pt \hbox{$\sim$} }}}

\newcommand{\bea}{\begin{eqnarray}}
\newcommand{\eea}{\end{eqnarray}}

\begin{document}

\baselineskip=15pt
\preprint{}

\title{Two Higgs doublet models augmented by a scalar colour octet}

\author{Li Cheng$^{1}$\footnote{Electronic address: lcheng@iastate.edu} and German Valencia$^{2}$\footnote{Electronic address: German.Valencia@monash.edu }}


\affiliation{$^{1}$ Department of Physics and Astronomy, Iowa State University, Ames, IA 50011.}

\affiliation{$^{2}$ School of Physics and Astronomy, Monash University, Melbourne, Australia.\footnote{On leave from Department of Physics, Iowa State University, Ames, IA 50011.}}

\date{\today}

\vskip 1cm
\begin{abstract}

The LHC is now studying in detail the couplings of the Higgs boson in order to determine if there is new physics. Many recent studies have examined the available fits to Higgs couplings from the perspective of constraining two Higgs doublet models (2HDM). In this paper we extend those studies to include constraints on the one loop couplings of the Higgs to gluons and photons. These couplings are particularly sensitive to the existence of new coloured particles that are hard to detect otherwise and we use them to constrain a 2HDM augmented  with a colour-octet scalar, a possibility motivated by minimal flavour violation. We first study theoretical constraints on  this model and then compare them with LHC measurements. 

\end{abstract}

\maketitle

\section{Introduction}

Following up on their discovery of the Higgs boson with mass near 125 GeV \cite{Aad:2012tfa,Chatrchyan:2012ufa}, the ATLAS and CMS collaborations continue the detailed study of its properties. For example, the Higgs couplings to top, bottom and tau have been measured to be in  agreement with the standard model (SM) although the errors are still large. Couplings to $WW$, $ZZ$ as well as effective one-loop couplings $h\to gg$ and $h\to \gamma\gamma$  are also well described by the SM \cite{atl-cms}.  
However, present day uncertainties still allow for a variety of new physics possibilities. For example, when compared with two Higgs doublet models (2HDM), these measurements constrain the parameter space but do not exclude the possibility of additional scalars below 1 TeV \cite{Celis:2013rcs,Krawczyk:2013gia,Barroso:2013zxa,Ferreira:2013qua,Dorsch:2013wja,Celis:2013ixa,Chang:2013ona,Dumont:2014wha,Ferreira:2014qda,Baglio:2014nea,Haber:2015pua}. Two Higgs doublet models can also be confronted with $h\to gg$ and $h\to \gamma\gamma$ fits and this comparison restricts the allowed parameter space.

Manohar and Wise (MW) \cite{Manohar:2006ga} introduced a model consisting of the SM augmented by a colour octet electroweak doublet of scalars. The addition was motivated by minimal flavour violation: assuming that the scalars transform trivially under the flavour group, only electroweak doublets which are colour singlets or octets are allowed. These coloured scalars are very weakly constrained by direct searches at LHC but they can affect the loop induced Higgs couplings by factors of two. The model has been constrained theoretically and also using the $h\to gg$ and $h\to \gamma\gamma$ fits  with comparable results, and there are many phenomenological studies in the literature \cite{Burgess:2009wm,Carpenter:2011yj,Enkhbat:2011qz,He:2011ti,Dobrescu:2011aa,Bai:2011aa,Arnold:2011ra,He:2011ws,Cacciapaglia:2012wb,Dorsner:2012pp,Kribs:2012kz,Reece:2012gi,Cao:2013wqa,He:2013tla,He:2013tia,Cheng:2015lsa,Buttazzo:2014bka,He:2014xla,Yue:2014tya,Kobakhidze:2014gqa,Cao:2015twy}. 

In this paper we combine these two extensions of the SM and consider a two Higgs doublet model with an additional scalar octet as in MW. The motivation for studying this model is that this is a simple extension of the SM that can satisfy minimal flavour violation. More complicated models exist that contain both of these ingredients \cite{Bertolini:2013vta,Perez:2016qbo}, but our approach here is purely phenomenological. Our main goal is to explore the one-loop effective couplings $h\to gg$ and $h\to \gamma\gamma$ of the SM-like Higgs in two Higgs doublet models in the presence of the additional scalar $S$ transforming as $(8,2,1/2)$ under the  SM gauge group $SU(3)_C\times SU(2)_L\times U(1)_Y$.

The model contains a large number of parameters that we first reduce by imposing standard theoretical constraints such as minimal flavour violation  \cite{Chivukula:1987py,D'Ambrosio:2002ex}, custodial symmetry \cite{Sikivie:1980hm,Pomarol:1993mu,Grzadkowski:2010dj}, and perturbative unitarity \cite{Lee:1977eg,Kanemura:1993hm, Horejsi:2005da,Ginzburg:2005dt,Grinstein:2015rtl,He:2013tla}. The question of vacuum stability \cite{Holthausen:2011aa,Degrassi:2012ry,vsrelated,EliasMiro:2012ay,Lebedev:2012zw,Rodejohann:2012px,Cheung:2012nb,Kannike:2012pe,Iso:2012jn,Bezrukov:2012sa,Tang:2013bz,Spencer-Smith:2014woa,Mohapatra:2014qva,Buttazzo:2013uya,Kobakhidze:2013pya,Grinstein:2013npa,Coleppa:2013dya,Craig:2015jba,Bernon:2015qea,
Dorsch:2016tab} is more complicated and will be discussed elsewhere.

\section{The model}

The model we discuss in this paper is an extension of the type I and type II two Higgs doublet models. In this extension we add a colour octet electroweak doublet of  scalars as in the MW \cite{Manohar:2006ga} extension of the SM. The scalar content is chosen to satisfy desirable properties: minimal flavour violation which naturally suppresses flavour changing neutral currents and custodial symmetry which naturally preserves the relation $\rho \approx 1$. As observed in Ref.~\cite{Manohar:2006ga}, the only possible extensions of the scalar sector that do not transform under the flavour group and that satisfy minimal flavour violation are electroweak doublets that are colour singlets or colour octets and this motivates our choice for this model. 

The scalar content of the model consists of two $SU(2)$ scalars ($\Phi_1, \Phi_2)$ and one colour-octet scalar $S$. The general potential for ($\Phi_1, \Phi_2)$ is well known from the literature \cite{Gunion:1989we,Branco:2011iw}. Our starting point will be more modest, consisting of the CP conserving, two Higgs doublet model with a discrete symmetry $\Phi_1\to -\Phi_1$ that is only violated softly by dimension two terms\footnote{This is more restrictive than MFV and we comment on this later on.}
\begin{eqnarray}
V\left( \Phi_1, \Phi_2 \right)  &=& m_{11}^2 \Phi_1^\dag \Phi_1 + m_{22}^2 \Phi_2^\dag \Phi_2 - m_{12}^2 \left( \Phi_1^\dag \Phi_2 + \Phi_2^\dag \Phi_1 \right) \nonumber \\
&+& \frac{\lambda_1}{2} \left( \Phi_1^\dag \Phi_1 \right)^2 + \frac{\lambda_2}{2} \left( \Phi_2^\dag \Phi_2 \right)^2 
+ \lambda_3 \left( \Phi_1^\dag \Phi_1 \right) \left( \Phi_2^\dag \Phi_2 \right)  \nonumber\\
&+& \lambda_4 \left( \Phi_1^\dag \Phi_2 \right) \left( \Phi_2^\dag \Phi_1 \right) + \frac{\lambda_5}{2} \left[ \left( \Phi_1^\dag \Phi_2 \right)^2 + \left( \Phi_2^\dag \Phi_1 \right)^2 \right].
\label{thdmV}
\end{eqnarray}

To this starting block we can add the most general, renormalizable potential that describes the couplings of the colour octet $S$ to the two colour singlets ($\Phi_1, \Phi_2)$ as well as the self interactions of the colour octet. This potential can be easily constructed by analogy with Ref.~\cite{Manohar:2006ga}, changing the notation for couplings to accommodate the standard use in Eq.~\ref{thdmV}. The octet self interactions do not change, but we use $\mu_{1-6}$ instead of $\lambda_{6-11}$ to label them,
\begin{eqnarray}
V(S) &=& 2m_S^2 {\rm Tr}S^{\dag i}S_i + \mu_1 {\rm Tr}S^{\dag i}S_i S^{\dag j}S_j + \mu_2 {\rm Tr}
S^{\dag i}S_j S^{\dag j}S_i + \mu_3 {\rm Tr} S^{\dag i}S_i {\rm Tr}S^{\dag j} S_j\nonumber\\
& +& \mu_4 {\rm Tr}S^{\dag i}S_j {\rm Tr}S^{\dag j}S_i + \mu_5 {\rm Tr}S_i S_j{\rm Tr}
S^{\dag i}S^{\dag j} + \mu_6 {\rm Tr}S_i S_j S^{\dag j}S^{\dag i}.
\label{sselfV}
\end{eqnarray}
The interactions between each one of the two colour singlets and the colour octet also follow Ref.~\cite{Manohar:2006ga} but using $\nu_{1-5}$ for $\Phi_1$ or $\omega_{1-5}$ for $\Phi_2$ in place of $\lambda_{1-5}$,
\begin{eqnarray}
V\left( \Phi_1, S \right)  &= &\nu_1 \Phi_1^{\dag i}\Phi_{1i}{\rm Tr}S^{\dag j}S_j + \nu_2 \Phi_1^{\dag i}\Phi_{1j}
{\rm Tr}S^{\dag j}S_i\nonumber\\
& +& \left( \nu_3 \Phi_1^{\dag i}\Phi_1^{\dag j}{\rm Tr}S_i S_j + \nu_4 \Phi_1^{\dag i}{\rm Tr}
S^{\dag j}S_j S_i + \nu_5 \Phi_1^{\dag i}{\rm Tr}S^{\dag j}S_i S_j + {\rm h.c.} \right) \nonumber \\
V\left( \Phi_2, S \right)  &= &\omega_1 \Phi_2^{\dag i}\Phi_{2i}{\rm Tr}S^{\dag j}S_j + \omega_2 \Phi_2^{\dag i}\Phi_{2j}
{\rm Tr}S^{\dag j}S_i\nonumber\\
& +& \left( \omega_3 \Phi_2^{\dag i}\Phi_2^{\dag j}{\rm Tr}S_i S_j + \omega_4 \Phi_2^{\dag i}{\rm Tr}
S^{\dag j}S_j S_i + \omega_5 \Phi_2^{\dag i}{\rm Tr}S^{\dag j}S_i S_j + {\rm h.c.} \right) 
\label{phisV}
\end{eqnarray}
Some of the couplings $\nu_{3,4,5}$ and $\omega_{3,4,5}$ can be complex and violate CP, but we will restrict our study to the CP conserving case. 
Finally, we have terms that involve both $\Phi_1$ and $\Phi_2$ as well as $S$,\footnote{Note that these terms are allowed by MFV but not by the discrete symmetry commonly used to restrict the 2HDM potential.}
\begin{eqnarray}
V_{N}\left( \Phi_1, \Phi_2, S \right) = \kappa_1 \Phi_1^{\dag i}\Phi_{2i}{\rm Tr}S^{\dag j}S_j +
\kappa_2 \Phi_1^{\dag i}\Phi_{2j}{\rm Tr}S^{\dag j}S_i + \kappa_3 \Phi_1^{\dag i}\Phi_2^{\dag j}{\rm Tr}S_j S_i
+ {\rm h.c.}
\label{3fields}
\end{eqnarray}
in all cases we have explicitly shown the  $SU(2)$ indices $i, j$, $S_i = T^A S_i^A$, and the trace is taken over colour indices.
The complete potential is thus,
\begin{equation}
V\left( \Phi_1, \Phi_2, S \right) = V\left( \Phi_1, \Phi_2 \right) + V (S) + V\left( \Phi_1, S \right) +
V\left( \Phi_2, S \right) + V_{N}\left( \Phi_1, \Phi_2, S \right).
\end{equation}

After symmetry breaking,  this potential implies the following relations between couplings and scalar masses
\begin{eqnarray}
m_{H^\pm}^2 &=& \frac{2 m_{12}^2}{\sin2\beta}-\frac{\lambda_4 + \lambda_5}{2}v^2,\quad\quad m_A^2 \, =\, \frac{2 m_{12}^2}{\sin2\beta} - \lambda_5 v^2,\nonumber \\
m_h^2 &=& \frac{2 m_{12}^2}{\sin2\beta} \cos^2 (\beta - \alpha)  + v^2 \left( \lambda_1 \sin^2 \alpha \cos^2 \beta + \lambda_2 \cos^2 \alpha \sin^2 \beta - \frac{\lambda_{345}}{2}\sin2\alpha \sin2\beta \right),\nonumber \\
m_H^2 &=& \frac{2 m_{12}^2}{\sin2\beta} \sin^2 (\beta - \alpha) + v^2 \left( \lambda_1 \cos^2 \alpha \cos^2 \beta + \lambda_2 \sin^2 \alpha \sin^2 \beta + \frac{\lambda_{345}}{2}\sin2\alpha \sin2\beta \right),\nonumber \\
m_{12}^2 &=& \frac{ v^2 \left[ \left(  \lambda_1 \cos^2 \beta - \lambda_2 \sin^2 \beta \right) \tan2\alpha - \frac{\lambda_{345}}{2}\sin2\beta \right]}{2 \tan2\alpha \cot2\beta - 1}.
\label{masses}
\end{eqnarray}
where $\lambda_{345}=\lambda_3+\lambda_4+\lambda_5$, and $v^2=v_1^2+v_2^2$ with $v_{1,2}$ the vevs of $\Phi_{1,2}$ respectively.  Similarly, for the colour octet sector we obtain 
\begin{eqnarray}
m_{{S^ \pm }}^2 &=& m_S^2 + \frac{v^2}{4}\left( \nu_1 \cos^2 \beta + \omega_1 \sin^2 \beta +
\kappa_1 \sin2\beta \right),\nonumber \\
m_{S_R^0}^2 &=& m_S^2 + \frac{v^2}{4} \left[ \left( \nu_1 + \nu_2 + 2 \nu_3 \right) \cos^2 \beta + \left( \omega_1 + \omega_2 + 2 \omega_3 \right) \sin^2 \beta \right. \nonumber\\
&+& \left. \left( \kappa_1 + \kappa_2 + \kappa_3 \right) \sin2\beta \right], \nonumber \\
m_{S_I^0}^2 &=& m_S^2 + \frac{v^2}{4} \left[ \left( \nu_1 + \nu_2 - 2 \nu_3 \right) \cos^2 \beta + \left( \omega_1 + \omega_2 - 2 \omega_3 \right) \sin^2 \beta \right. \nonumber\\
&+& \left.  \left( \kappa_1 + \kappa_2 - \kappa_3 \right) \sin2\beta \right].
\label{masses2}
\end{eqnarray}

The Yukawa couplings in this model consist of two types of terms that we can write as
\begin{equation}
L_Y = L_{Y1}\left( \Phi_1, \Phi_2 \right) + L_{Y2}\left( S \right)
\end{equation}
corresponding to the usual two Higgs doublet model couplings plus the interactions of the fermions with the colour octet. In the flavour eigenstate basis, they are
\begin{eqnarray}
&L_{Y1}\left( \Phi_1, \Phi_2 \right) =  - {\left( g_1^D \right)^\alpha}_\beta {\bar D}_{R, \alpha }\Phi_1^\dag Q_L^\beta -
{\left( g_1^U \right)^\alpha}_\beta {\bar U}_{R, \alpha}{\tilde \Phi}_1^\dag Q_L^\beta \nonumber\\
&\qquad\qquad\qquad\;\: - {\left( g_2^D \right)^\alpha}_\beta {\bar D}_{R, \alpha } \Phi_2^\dag Q_L^\beta  - {\left( g_2^U
\right)^\alpha}_\beta {\bar U}_{R, \alpha}{\tilde \Phi}_2^\dag Q_L^\beta + {\rm h.c.},\nonumber \\
&L_{Y2}(S) =  - {\left( g_3^D \right)^\alpha}_\beta {\bar D}_{R, \alpha}S^\dag Q_L^\beta  -
{\left( g_3^U \right)^\alpha}_\beta {\bar U}_{R, \alpha }{\tilde S}^\dag Q_L^\beta  + {\rm h.c.}
\label{yukawas}
\end{eqnarray}
where we have defined as usual ${\tilde H}_i = \varepsilon_{ij} H_j^*$
 for all three scalar doublets $H=\Phi_{1,2},S$, $S = T^A S^A$, and $\alpha, \beta$ are flavour indices.

\subsection{Minimal flavour Violation}

To suppress flavour changing neutral currents in two Higgs doublet models, it is conventional to introduce discrete symmetries. For the Type I model, $g_1^{D,U} = 0$, while in the Type II model, $g_1^U = g_2^D = 0$. In the Yukawa terms, the type I model can be enforced with the discrete symmetry $\phi_1 \to  - \phi_1$, whereas the type II model can be enforced with the discrete symmetry $\phi_1 \to - \phi_1$, $d_R \to  - d_R$ \cite{Gunion:1989we}. We will instead follow Ref.~\cite{Manohar:2006ga} and enforce MFV, requiring  that there be only two flavour symmetry breaking matrices $G^U$ transforming as $(3_U,\bar{3}_Q)$ under the flavour group and $G^D$ transforming as $(3_D,\bar{3}_Q)$ under the flavour group. The matrices appearing in Eq.~\ref{yukawas} must satisfy
\begin{eqnarray}
g_1^D=\eta_1^D G^D,\, g_2^D=\eta_2^D G^D,\, g_3^D=\eta_3^D G^D \nonumber\\
g_1^U=\eta_1^U G^U,\, g_2^U=\eta_2^U G^U,\, g_3^U=\eta_3^U G^U.
\end{eqnarray}
where $\eta_i^{D, U}$, $i = 1, 2, 3$, are complex scalars. The two types of two Higgs doublet model under consideration are then defined by 
\begin{itemize}
\item Type I: $\eta_1^D=\eta_1^U =0$ 
\item Type II: $\eta_1^U=\eta_2^D=0$
\end{itemize}
instead of the usual discrete symmetries. 

Requiring MFV instead of a discrete symmetry to define the models allows quartic terms in the scalar potential that are odd in either of the doublets. This justifies including the terms with coefficients $\nu_{4,5}$, $\omega_{4,5}$ and $\kappa_{1,2,3}$ in Eqs.~\ref{phisV}~and~\ref{3fields}.  One should note that in general, this also allows the additional terms in Eq.~\ref{thdmV},
\begin{eqnarray}
V^\prime(\Phi_1,\Phi_2) = \lambda_6 \left( \Phi_1^\dag \Phi_1 \right)\left( \Phi_1^\dag \Phi_2 \right)+\lambda_7 \left( \Phi_2^\dag \Phi_2 \right)\left( \Phi_1^\dag \Phi_2 \right) +{\rm h.c.}.
\end{eqnarray}
We will not include these two terms in our numerical studies for ease in comparing with the usual definitions of these two types of 2HDM, and because our main new ingredient is the colour octet sector.

\subsection{Custodial symmetry}

To impose custodial symmetry conveniently, we follow the matrix formulation of Ref.~\cite{Pomarol:1993mu} in which the scalar doublets are written as follows,
\begin{eqnarray}
&{M_{ab}} = \left( {{{\tilde \Phi }_a}, {\Phi _b}} \right) =
\begin{pmatrix}
{\phi _a^{0*}}&{\phi _b^ + }\\
{ - \phi _a^ - }&{\phi _b^0}
\end{pmatrix},\ 
a, b = 1,2,\\
&{{\cal S}^A} = \left( {{{\tilde S}^A}, {S^A}} \right) =
\begin{pmatrix}
{{S^{A0*}}}&{{S^{A + }}}\\
{ - {S^{A - }}}&{{S^{A0}}}
\end{pmatrix},
\end{eqnarray}
and the custodial symmetry is imposed by writing the scalar potential directly in terms of $O(4)$ invariants such as $\Phi _1^{\dag i}{\Phi _{2i}}{S^{\dag j}}{S_j} \to {\rm{Tr}}\left( {M_{11}^\dag
{M_{22}}} \right){\rm{Tr}}\left( {{{\cal S}^\dag }{\cal S}} \right)$.

There are two methods proposed in the literature, 
\begin{itemize}
\item {\bf Case 1.}
Construction using only ${M_{11}}$ and ${M_{22}}$. This yields the following constraints on the couplings of Eqs.~\ref{thdmV}-\ref{3fields}: all the $\lambda_i$ are real and 
\begin{eqnarray}
{\kappa_2} = {\kappa _3},\ 2{\nu _3} = {\nu _2},\ {\nu _4} = \nu _5^*,\ 2{\omega _3} = {\omega_2},\ {\omega _4} = \omega _5^*,\ \lambda_4 = \lambda_5.
\label{met1}
\end{eqnarray}

\item {\bf Case 2.}
Construction using only ${M_{12}}$ yielding instead the constraints
\begin{eqnarray}
&&{\nu_2}={\omega_2}={\kappa_3} = {\kappa _3^\star},\  \kappa_2=2\nu_2,\  \nu_3=\omega_3^\star,\ \nonumber \\
 &&\lambda_6 = \lambda_7,\ \lambda_1=\lambda_2=\lambda_3,\ m_{11}^2=m_{22}^2.
 \label{met2}
\end{eqnarray}
For the vacuum to be invariant as well one needs $v_1^\star=v_2$.
\end{itemize}

An immediate consequence of custodial symmetry is that  ${\rm{\Delta}} \rho = 0$ holds. The change induced  in ${\rm{\Delta}} \rho$ by the colour octet scalars is \cite{Manohar:2006ga}, 
\begin{equation}
{\rm{\Delta}} \rho \propto \left(v_1^2 \nu_2 + v_2^2 \omega_2 +2v_1 v_2 \kappa_2 \right)^2
- \left( 2v_1^2 \nu_3 + 2 v_2^2 \omega_3 + 2 v_1 v_2 \kappa_3 \right)^2.
\end{equation}
Upon substitution of Eqs.~\ref{met1}~and~\ref{met2} we find both sets of constraints result in  ${\rm{\Delta}} \rho = 0$ as expected.

As is known, both cases also in mass degeneracies $m_{H^\pm}=m_A$ and from Eqs.~\ref{masses},\ref{masses2} they also result in $m_{S^\pm}=m_{S_I^0}$. The constrain $v_1^\star=v_2$ is too restrictive so we will only use the first method, Eq.~\ref{met1} for our numerical study.

It has been pointed out before that it is also possible to satisfy ${\rm{\Delta}} \rho = 0$ with $m_{H^\pm}=m_H$ \cite{Gerard:2007kn,Cervero:2012cx} and with $m_{S^\pm}=m_{S_R^0}$ \cite{Burgess:2009wm}, and that this follows from `twisted' custodial symmetry.

\section{Unitarity and Stability Constraints}

In this section we consider high energy two-to-two scalar scattering to constrain the strength of the self interactions with the requirement of perturbative unitarity. The potential is renormalizable and the tree-level scattering amplitudes approach a constant value at high energy proportional to the quartic couplings. Perturbative unitarity then constrains their size as it does for the Higgs boson mass \cite{Lee:1977eg}. These constraints have been previously applied to two Higgs doublet models \cite{Kanemura:1993hm, Horejsi:2005da,Ginzburg:2005dt,Grinstein:2015rtl}, and to the Manohar-Wise model \cite{He:2013tla}. We extend them here to the combined model as described in the previous section, considering only the neutral, colour singlet  amplitudes. We begin by defining the two particle state basis for the calculation of amplitudes,
\begin{eqnarray}
&\left| {{A_a}} \right\rangle  = \frac{1}{{2\sqrt 2 }}\left| {2\phi _a^ + \phi
_a^ -  + {\rho _a}{\rho _a} + {\eta _a}{\eta _a}} \right\rangle,\ 
\left| {{B_a}} \right\rangle  = \frac{1}{{2\sqrt 2 }}\left| {2\phi _a^ + \phi
_a^ -  - {\rho _a}{\rho _a} - {\eta _a}{\eta _a}} \right\rangle, \nonumber \\
&\left| {{C_a}} \right\rangle  = \frac{1}{2}\left| {{\rho _a}{\rho _a} - {\eta_a}{\eta _a}} \right\rangle,\ 
\left| {{D_a}} \right\rangle  = \left| {{\rho _a}{\eta _a}} \right\rangle,\nonumber \\
&\left| {{E_1}} \right\rangle  = \frac{1}{{\sqrt 2 {\rm{i}}}}\left| {\phi _1^ +
\phi _2^ -  - \phi _2^ + \phi _1^ - } \right\rangle,\ 
\left| {{E_2}} \right\rangle  = \frac{1}{{\sqrt 2 }}\left| {{\rho _1}{\eta _2} -
{\rho _2}{\eta _1}} \right\rangle, \nonumber \\
&\left| {{F_ + }} \right\rangle  = \frac{1}{2}\left| {\phi _1^ + \phi _2^ -  +
\phi _2^ + \phi _1^ -  + {\rho _1}{\rho _2} + {\eta _1}{\eta _2}} \right\rangle,\ 
\left| {{F_ - }} \right\rangle  = \frac{1}{2}\left| {\phi _1^ + \phi _2^ -  +
\phi _2^ + \phi _1^ -  - {\rho _1}{\rho _2} - {\eta _1}{\eta _2}} \right\rangle, \nonumber \\
&\left| {{F_1}} \right\rangle  = \frac{1}{{\sqrt 2 }}\left| {{\rho _1}{\rho _2}
- {\eta _1}{\eta _2}} \right\rangle,\ 
\left| {{F_2}} \right\rangle  = \frac{1}{{\sqrt 2 }}\left| {{\rho _1}{\eta _2} +
{\rho _2}{\eta _1}} \right\rangle,\nonumber \\
&\left| {{S_1}} \right\rangle  = \frac{1}{8}\left| {2{S^{A + }}{S^{A - }} +
S_R^{A0}S_R^{A0} + S_I^{A0}S_I^{A0}} \right\rangle,\ 
\left| {{S_2}} \right\rangle  = \frac{1}{8}\left| {2{S^{A + }}{S^{A - }} -
S_R^{A0}S_R^{A0} - S_I^{A0}S_I^{A0}} \right\rangle, \nonumber \\
&\left| {{S_3}} \right\rangle  = \frac{1}{{4\sqrt 2 }}\left| {S_R^{A0}S_R^{A0} -
S_I^{A0}S_I^{A0}} \right\rangle,\ 
\left| {{S_4}} \right\rangle  = \frac{1}{{2\sqrt 2 }}\left| {S_R^{A0}S_I^{A0}}\right\rangle.
\end{eqnarray}
The unitarity constraints for the 2HDM without the coloured scalars are known from Ref.~\cite{Kanemura:1993hm,Ginzburg:2005dt}. The two-to-two scattering matrix is a $14\times 14$ matrix that can be diagonalized exactly producing the following eigenvalues (which we have simplified by setting $\lambda_5=\lambda_4$ as per custodial symmetry),
\begin{eqnarray}
\frac{3\left( \lambda_1 + \lambda_2 \right) \pm \sqrt{9\left( \lambda_1 - \lambda_2 \right)^2 + 4\left( 2\lambda_3 + \lambda_4 \right)^2}} 2,\nonumber \\
\frac{\left( \lambda_1 + \lambda_2 \right) \pm \sqrt{\left( \lambda_1 - \lambda_2 \right)^2
+ 4\lambda_4^2}} 2\; [3], \nonumber \\
(\lambda_3-\lambda_4)\; [2],\;  (\lambda_3+\lambda_4)\; [3],\;  (\lambda_3+5\lambda_4),\;
\label{unicon}
\end{eqnarray}
and have used the numbers in square brackets to denote the degeneracy of each particular eigenvalue. Unitarity constraints are obtained from the  $J=0$ partial waves of these two-to-two scattering amplitudes, by requiring that $|a_0|\leq1/2$. This is equivalent to requiring that the largest eigenvalue in Eq.~\ref{unicon} be less than $8\pi$.

In addition to the unitarity constraint, we also impose the known conditions for having a positive definite Higgs potential with a $Z_2$ symmetry \cite{Deshpande:1977rw},
\begin{equation}
\lambda_1 > 0, \quad
\lambda_2 > 0, \quad
\lambda_3 > - \sqrt{\lambda_1 \lambda_2},\quad
\lambda_3 + \lambda_4 \pm \lambda_5 > - \sqrt{\lambda_1 \lambda_2}.
\label{stabvac}
\end{equation}
For phenomenological studies one prefers to control the scalar masses instead of the $\lambda_i$ couplings as input parameters via the relations Eq.~\ref{masses}. We will always identify the lightest neutral scalar $h$ with the 125.6~GeV state found at LHC \cite{Aad:2012tfa,Chatrchyan:2012ufa}. The other masses will be allowed to vary in ranges discussed later on, but we will always use $\lambda's$ that ensure all the squared masses are positive and larger than around $(400~{\rm GeV})^2$. 

When we add the colour octet, the two-to-two scattering matrix becomes an $18\times 18$ matrix which we diagonalize numerically. 
Unitarity constraints are obtained again from the $J=0$ partial wave as in the case of the 2HDM. 
Approximate results in the custodial symmetry limit from $4\times 4$ submatrices are,
\begin{subequations}
\begin{align}
&\left| \lambda_1 \right|,\left| \lambda_2 \right| \le \frac{8\pi}{3},\quad
\left| \lambda_3 \right| \le 4\pi,\quad \left| \lambda_4 \right|, \left| \lambda_5 \right| \le \frac{8\pi}{5},\\
&\left| \nu_1 \right|,\left| \nu_3 \right|, \left| \omega_1 \right|, \left| \omega_3 \right| \le 2\sqrt{2}\pi,\quad
\left| \nu_2 \right|,\left| \omega_2 \right| \le 4\sqrt{2}\pi,\\
&\left| \kappa_1 \right| \le 2\pi,\quad \left| \kappa_2 \right|, \left| \kappa_3 \right| \le 4\pi.
\label{approxlim}
\end{align}
\end{subequations}
The couplings that affect only octet self-interactions at tree level, those in $V(S)$ Eq.~\ref{sselfV}, have identical constraints as already found in Ref.~\cite{He:2013tla}. In particular Eq.~3.9 of that reference (translated to the notation of this paper)
\begin{equation}
\left| 17 \mu_3 +13 \mu_4 +13 \mu_6 \right|\leq 16 \pi
\label{mu1con}
\end{equation}
is reproduced in our numerical diagonalization of the $18\times 18$ matrix. Additional constraints obtained in Ref.~\cite{He:2013tla} by studying unitarity in the colour octet channel are imposed on our entries and we quote them here for convenience,
\begin{eqnarray}
|\nu_4+\nu_5|\lsim \frac{32\pi}{\sqrt{15}}, \quad
|\omega_4+\omega_5| \lsim  \frac{32\pi}{\sqrt{15}}, \quad
|2 \mu_3+10 \mu_4+7\mu_6|\leq 32\pi.
\label{mu2con}
\end{eqnarray}
We illustrate the constraints resulting from perturbative unitarity in several figures to be described below.

\section{Existing LHC constraints}

\subsection{Tree-level Higgs decay}

The tree-level Higgs couplings to fermion pairs, in particular $t\bar{t}$, $b\bar{b}$ and $\tau^+\tau^-$ as well as the couplings to $W$ and $Z$ already constrain the parameter space of the 2HDM requiring it to be close to the SM. Allowed regions of parameter space under different scenarios have been presented recently for example in Ref.~\cite{Barroso:2013zxa,Ferreira:2013qua,Haber:2015pua,Craig:2015jba,Dorsch:2016tab} and we do not repeat this exercise. The reader interested in the results of that global fit is referred to Figure~1 in Ref.~\cite{Dorsch:2016tab}, for example.

\begin{figure}[thb]
\includegraphics[width=.9\textwidth]{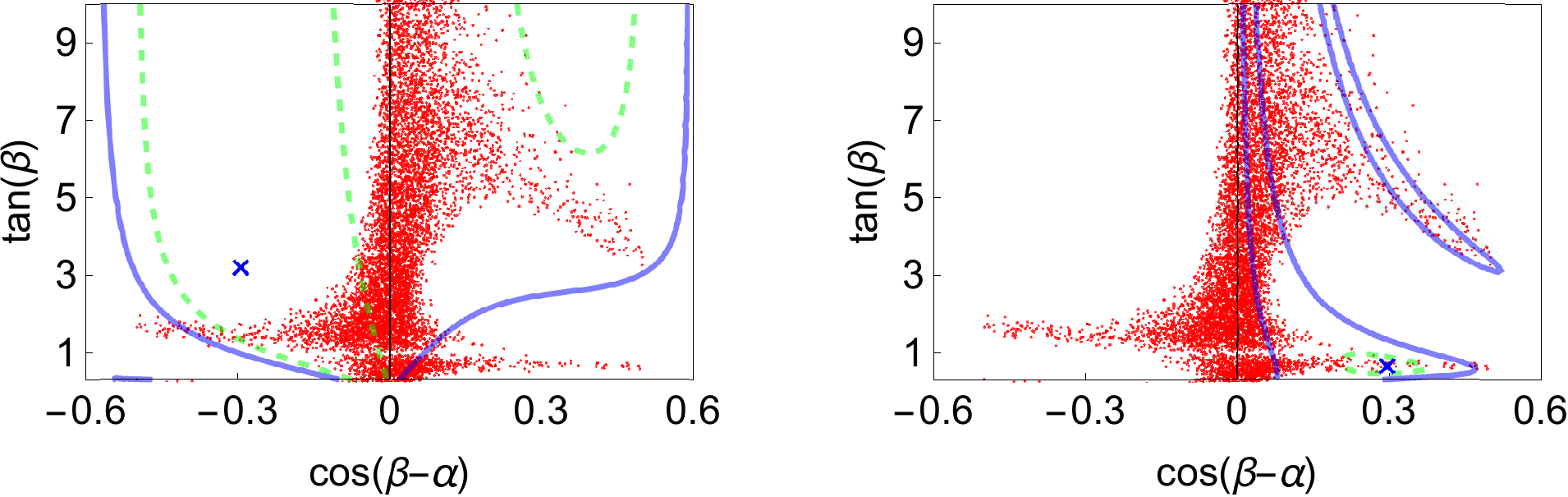}
\caption{
\label{tree-con} $\chi^2$ fit to the couplings in Eq.~\ref{atcmsfit} shown in the $\cos(\beta-\alpha)-\tan\beta$ plane. In the left panel we have the 2HDM-I and in the right panel we have the 2HDM-II. In both cases the blue cross marks the best fit and the blue contour encloses the region allowed at $95\%$ confidence level. The dashed green shows the $68\%$ c.l. region.  Superimposed is the red dotted area corresponding to points allowed by tree-level unitarity.}
\end{figure}

There are a few relevant comments to be made that are not apparent from the global fit. To this end we consider the results of the seven parameter fit to the Higgs couplings as per the ATLAS-CMS combination of data. We further consider their second scenario, in which contributions from BSM particles are allowed both in the loops and in the Higgs decay but $\kappa_V\leq1$ is assumed. Those results, as listed on Table~14 of  \cite{atl-cms} are:
\begin{eqnarray}
\kappa_b = 0.57^{+0.16}_{-0.16},\, 
\kappa_\tau = 0.87^{+0.12}_{-0.11},\,
\kappa_t = 1.42^{+0.23}_{-0.22},\nonumber\\
\kappa_Z=1.00_{-0.08},\, \kappa_W=0.90^{+0.09}_{-0.09}.
\label{atcmsfit}
\label{usedvalues}
\end{eqnarray}
Recalling that in 2HDM-I
\begin{eqnarray}
\kappa_t=\kappa_b=\kappa_\tau=\frac{\cos(\beta-\alpha)}{\tan\beta}+\sin(\beta-\alpha)
\end{eqnarray}
one sees that the $b$ and $t$ couplings to the Higgs from Eq.~\ref{atcmsfit}  are in tension within the 2HDM-1, being a bit more than $3\sigma$ away if one adds the two errors in quadrature. 
To connect with the usual plot presented in the literature \cite{Haber:2015pua,Craig:2015jba,Dorsch:2016tab}, we can do a simple fit to the 5 couplings in Eq.\ref{atcmsfit}, which we show in 
Figure~\ref{tree-con}. The left panel illustrates the same point as the best fit is closer to $\kappa_b$ and so is the $68\%$ c.l. region enclosing the best fit point. The second dashed-green region is closer to $\kappa_t$ and one needs to go to a 95\% c.l. to obtain a connected region which covers most of the parameter space. The addition of the colour octet cannot help address this problem as it does not affect the fermion Yukawa couplings at tree-level.

On the right panel we repeat the comparison for the type-II 2HDM. In this case there is a much smaller allowed region of parameter space but the goodness of the fit (as measured by $\chi^2_{\rm min}$) is better than that for 2HDM-I. The blue contour is similar, but not identical, to that obtained in the literature from a direct global fit to LHC measurements. The slight shift of this region towards larger values of $\cos(\beta-\alpha)$ is due to the small value of $\kappa_b$ and its small error in Eq.~\ref{atcmsfit}.

The values of $\kappa_Z=1.00_{-0.08}$ and $\kappa_W=0.90\pm0.09$  in Eq.~\ref{atcmsfit} prefer the region $\cos(\beta-\alpha)$ near one, the so called alignment limit. 
In addition there are constraints from the non-observation of the additional Higgs bosons that are shown in Ref.~\cite{Haber:2015pua}, for example, and that we do not reproduce here. The constraints shown Figure~\ref{tree-con} are not affected by the additional coloured scalars and should be identical to those obtained in the 2HDM if the same constraints are used. For this reason, they are not directly the concern of this paper.

\subsection{Direct bounds on the colour octet}

One would expect that the LHC can place stringent constraints on the existence of the additional colour scalars from their non-observation. It turns out however that the existing bounds are not very restrictive for this model, depending on the values of the couplings in the scalar potential the masses. The main reason is that the cross-sections for production of one or two such scalars are below current LHC sensitivity as can be ascertained by a quick glance at theoretical predictions \cite{Gresham:2007ri,Arnold:2011ra} compared to those for coloured scalars that are currently constrained \cite{Han:2010rf} and vis-a-vis LHC results \cite{Chatrchyan:2012yca,Khachatryan:2015dcf}. Indirect constraints allow masses as low as $\sim 100$~GeV \cite{Burgess:2009wm}.

The most important decays of the neutral scalars for example, would be into two jets or a $ t\bar{t}$ pair. CMS limits on a colour-octet scalar $S^0$ from dijet final state quote $M_S < 3.1$~TeV \cite{Khachatryan:2015dcf}. However, this is a gross overestimate for the MW model where the $S^0$ production cross-section is a few thousand times smaller than the model used by CMS. Similarly, bounds on $Z^\prime$ resonances decaying to  $ t\bar{t}$ pairs \cite{Chatrchyan:2012yca} can be interpreted as posing no significant constraint for these scalars where $\sigma_S B(S\to t\bar{t}) \sim 50-100$~fb since their best sensitivity is to $\sigma_S B(S\to t\bar{t}) \gsim 200$~fb for the mass range studied (up to 2 TeV for narrow resonances and 3 TeV for wide resonances).

As already mentioned in Ref.~\cite{Manohar:2006ga,Gresham:2007ri} the cross sections for producing pairs of coloured scalars are larger than those for single scalar production for much of the parameter space. In this case the relevant constraints would arise from searches for dijet pairs and four top-quarks. Again the relevant quantity  $\sigma_S Br^2$ for this model is measured in fb whereas the published constraints are above this. Nonetheless, the dijet pair channel appears to be the most promising one to constrain this model and a detailed study will be forthcoming.

For our numerical study we will use two examples, one in which $M_{S^\pm}$ is set at 1~TeV and another one at 800~GeV. The couplings in the potential affecting Eq.~\ref{masses2} are constrained so that $725\leq M_{S^0_R}\leq 1200$~GeV, and the custodial symmetry will ensure that  $M_{S^0_I}=M_{S^\pm}$.

\section{One-loop decays of neutral colour-singlet scalars to $gg$ and $\gamma\gamma$}

Finally we discuss the loop induced Higgs couplings where the colour-octet can play its most important role. 
Fits to the LHC Higgs data already exist in the literature and we use Ref.~\cite{Giardino:2013bma} for our discussion. It is standard to parameterize the one-loop results with effective operators for  $hgg$ and $h\gamma\gamma$
\begin{equation}
\mathcal{L}_{\textrm{eff}}=c_g \frac{\alpha_s}{12\pi v}h G^a_{\mu\nu}G^{a\mu\nu}+c_\gamma \frac{\alpha}{\pi v}h F_{\mu\nu}F^{\mu\nu}.
\end{equation}
A general parametrization for couplings to the Higgs of different kinds of new particles such as a complex scalar  $S$, a Dirac fermion $f$, and a charged and colourless vector $V_\mu$ are
\begin{equation}
\mathcal{L} = -  c_s \frac{2 M_S^2}{v}  h  S^\dagger S  - c_f \frac{M_f}{v} h \bar f f  +  c_V \frac{2 M_V^2 }{v} h V_\mu^\dagger  V^\mu.
\label{efftree}
\end{equation}
They contribute to the effective Higgs coupling to gluons and to photons at one-loop as \cite{Ellis:1975ap,Ioffe:1976sd,Shifman:1979eb,Djouadi:2005gi}
\begin{eqnarray}
\delta c_g   &=&  \frac{3C_2(r_s)}{ 2}  c_s A_s(\tau_s) +  \frac{3C_2(r_f)}{2} c_f A_f(\tau_f),\\
\delta c_\gamma    &=&   \frac{ N(r_s) Q_s^2 }{8}  c_s A_s(\tau_s) + \frac{ N(r_f) Q_f^2}{8} c_f A_f(\tau_f)  -  \frac{Q_V^2}{8} c_V A_V(\tau_V),
\end{eqnarray}
where $\delta c_i = c_i - c_{i,\rm SM} $,  $C_2(r)$ is the quadratic Casimir of the colour representation $r$, and $N(r)$ is the number of colours of the representation $r$. $A_i$ ($i = S, f, V$, standing for scalar boson loop, fermion loop and vector boson loop, respectively) are loop functions,
\begin{subequations}
\begin{align}
&A_S(\tau) = \frac{3}{\tau^2}(f(\tau ) - \tau ),\\
&A_f(\tau) =  \frac{3}{2\tau^2}((\tau  - 1)f(\tau ) + \tau ),\\
&A_V(\tau) = \frac{1}{7\tau^2}(3(2\tau  - 1)f(\tau ) + 3\tau  + 2\tau^2),
\end{align}
\end{subequations}
with
\begin{align*}
&f(\tau) = \begin{cases}
{\arcsin ^2}\sqrt \tau , &\tau \le 1, \\
- \frac{1}{4}\left( \ln \frac{\eta^+}{\eta^-} - {\rm{i}}\pi \right)^2, &\tau > 1,
\end{cases} \\
&\eta ^ \pm = 1 \pm \sqrt {1 - 1/\tau}.
\end{align*}

In terms of these general results and using
\begin{equation}
r_i = \frac{m_h^2}{4m_i^2},  \,\,R_i=\frac{m_H^2}{4m_i^2}, 
\end{equation}
we can write the effective one loop couplings. We begin quoting, for completeness, 
the amplitudes for these two processes within the SM  \cite{Djouadi:2005gi},
\begin{eqnarray}
M(h\to gg)_{SM} &=&  A_f(r_t)  \nonumber \\
M(h\to\gamma\gamma)_{SM} &=& \frac{2 }{9 }A_f(r_t)  - \frac{7 }{8} A_V(r_W) 
\end{eqnarray}
Similarly the one-loop $\gamma\gamma$ and $gg$ couplings  for the 2HDM neutral scalars are given by 
\begin{eqnarray}
M(h\to gg)_{2HDM} &=& \frac{\cos\alpha  }{\sin\beta}  A_f(r_t)  + \frac{\cos\alpha  }{\sin\beta} 
     A_f(r_b) t_1 - \frac{\sin\alpha  }{\cos\beta } A_f(r_b) t_2 \nonumber\\
M(H\to gg)_{2HDM} &=& \frac{\sin\alpha  }{\sin\beta}  A_f(R_t)  + \frac{\sin\alpha  }{\sin\beta }
     A_f(R_b) t_1 + \frac{\cos\alpha  }{\cos\beta } A_f(R_b) t_2 \nonumber\\
M(h\to\gamma\gamma)_{2HDM} &= &
  \frac{2 }{9}\frac{\cos\alpha  }{\sin\beta}  A_f(r_t)  + 
   \frac{1 }{18}\frac{\cos\alpha  }{\sin\beta}  A_f(r_b) t_1 - 
  \frac{ 1 }{18}\frac{\sin\alpha  }{\cos\beta}  A_f(r_b) t_2\nonumber \\ &-& 
  \frac{ 7 }{8}\sin(\beta - \alpha) A_V(r_W)  +\frac{ 1 }{48}g_{hH^\pm} A_s(r_{H^+})  \nonumber\\
M(H\to\gamma\gamma)_{2HDM} &= &
  \frac{2 }{9}\frac{\sin\alpha  }{\sin\beta}  A_f(R_t)  + 
  \frac{ 1 }{18}\frac{\sin\alpha  }{\sin\beta}  A_f(R_b) t_1 + 
   \frac{1 }{18}\frac{\cos\alpha  }{\cos\beta}  A_f(R_b) t_2 \nonumber \\ &-& 
   \frac{7 }{8}\cos(\beta - \alpha) A_V(R_W)  + \frac{1 }{48}g_{HH^\pm} A_s(R_{H^+})  
\end{eqnarray}
where $t_1=1,t_2=0$ for Type-I and  $t_1=0,t_2=1$ for Type-II and 
\begin{eqnarray}
g_{hH^\pm} &= &\frac{v^2}{m_{H^\pm} ^2}\left(-\lambda_1\sin\alpha \sin^2\beta \cos\beta  + 
     \lambda_2\cos\alpha \sin\beta \cos^2\beta  \right. \nonumber \\ &+& \left.
     \lambda_3(\cos\alpha \sin^3\beta  - \sin\alpha \cos^3\beta )
    - 2\lambda_4\cos(\alpha + \beta)\sin\beta \cos\beta\right ) \nonumber\\
g_{HH^\pm} &= & \frac{v^2 }
   {m_{H^\pm} ^2}\left(\lambda_1\cos\alpha \sin^2\beta \cos\beta  + 
     \lambda_2\sin\alpha \sin\beta \cos^2\beta  \right. \nonumber \\ &+& \left.
     \lambda_3(\cos\alpha \cos^3\beta  + \sin\alpha \sin^3\beta )-
     2\lambda_4\sin(\alpha + \beta)\sin\beta \cos\beta \right) 
        \end{eqnarray}
The top-quark and $W$-boson contributions to $M(h\to gg)$ and $M(h\to \gamma\gamma)$ in the above expressions for the 2HDM, reduce to the SM in the limit $\beta-\alpha = \frac \pi 2$. The colour octet scalars  contribute the additional terms 
\begin{eqnarray}
M(h\to \gamma\gamma)_{S} &=&   \frac{1 }{3}c^\pm A_s(r_{S^\pm})   \nonumber\\
M(H \to \gamma\gamma)_{S} &=&   \frac{1 }{3}C^\pm A_s(R_{S^\pm}) \nonumber\\
  M (h\to gg)_{S}  &=& \frac{3 }{2}c^\pm A_s(r_{S^\pm})  + 
  \frac{ 3 }{4}c^rA_s(r_{S_R})  +\frac{ 3 }{4}c^i A_s(r_{S_I})  \nonumber\\
M(H\to gg)_{S} &=& \frac{3 }{2}C^\pm A_s(R_{S^\pm} ) + 
   \frac{3 }{4}C^r A_s(R_{S_R})  + \frac{3 }{4}C^i A_s(R_{S_I})  
\label{extraloop}
\end{eqnarray}
where 
\begin{eqnarray}
     c^\pm &=&  \frac{v^2 }{4M_{S^\pm}^2}(-\nu_1\sin\alpha \cos\beta  + \
\omega_1\cos\alpha \sin\beta  + \kappa_1\cos(\alpha + \beta))
  \nonumber \\
c^r &=&\frac{v^2 }{4M_{S_R}^2} (-(\nu_1 + 2\nu_2)\sin\alpha \cos\beta  + (\omega_1 + 
        2\omega_2)\cos\alpha \sin\beta  + (\kappa_1 + 
        2\kappa_2)\cos(\alpha + \beta)) \nonumber\\
C^\pm &=&\frac{v^2 }{4M_{S^\pm}^2 } (\nu_1\cos\alpha \cos\beta  + \omega_1\sin\alpha \sin\
\beta  + \kappa_1\sin(\alpha + \beta)) \nonumber\\
C^r &=& \frac{v^2 }{4M_{S_R}^2 }((\nu_1 + 2\nu_2)\cos\alpha \cos\beta  + (\omega_1 + 
        2\omega_2)\sin\alpha \sin\beta  + (\kappa_1 + 
        2\kappa_2)\sin(\alpha + \beta)) \nonumber\\
C^i &=& C^\pm, \quad\quad c^i \,=\, c^\pm
\end{eqnarray}
where we have shown our results in the custodial $SU(2)$ limit, and the total contributions for the models in this work are $M_{2HDM}+M_{S}$.

\section{Numerical study}

\begin{figure}[htb]
\includegraphics[width=.9\textwidth]{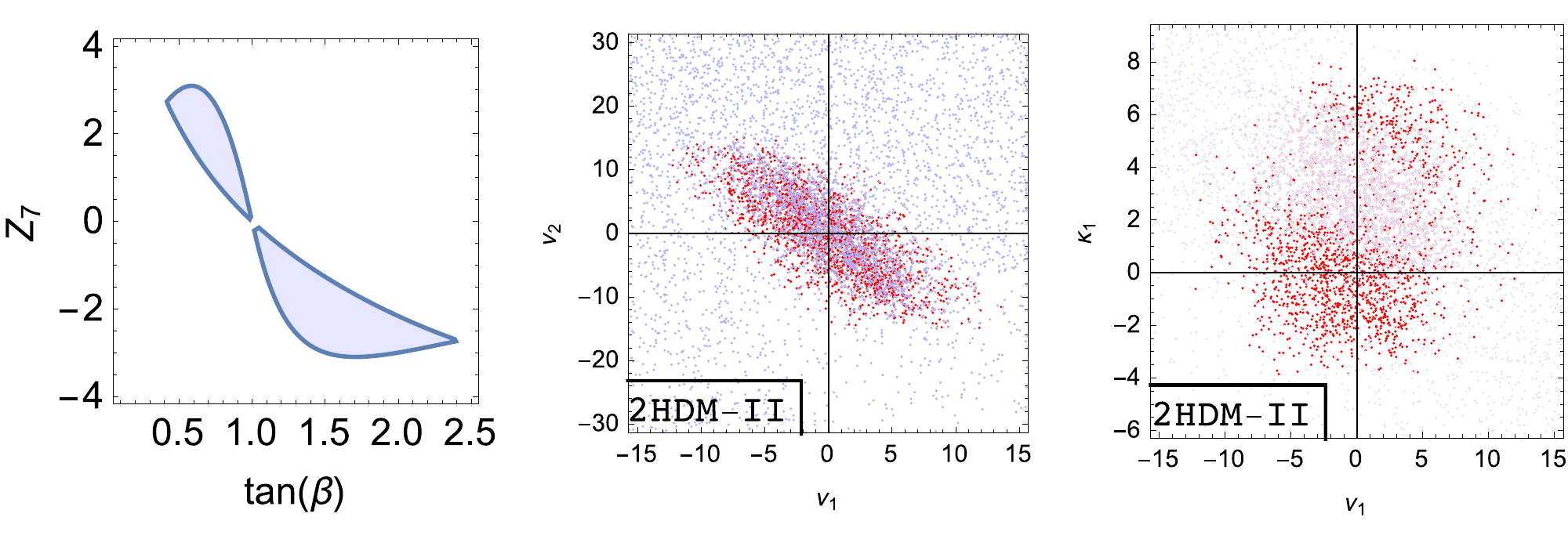}
\caption{Left panel: allowed $\tan\beta-Z_7$ parameter space for the example discussed in the text. Center panel: unitarity constraints in $\nu_1-\nu_2$ for the same example (red points) and (blue points) allowed by $h\to \gamma\gamma$ and $h\to gg$ at $1\sigma$. Right panel: unitarity constraints in $\nu_1-\kappa_1$ for the same example (red points) and (blue points) allowed by $h\to \gamma\gamma$ and $h\to gg$ at $1\sigma$.
}
\label{example} 
\end{figure}

The model contains a large number of free parameters so we begin by presenting numbers for special values of masses to get a simple picture. We assume the lighter neutral CP-even Higgs $h$ is the one discovered at LHC, and then compare the branching ratios to $gg$ and $\gamma\gamma$ to the fit of Ref.~\cite{Giardino:2013bma}. We first set $\beta-\alpha = \frac \pi 2$,  $m_{H^\pm} = 600\ {\rm GeV}$, $m_A = 500\ {\rm GeV}$,  $m_{S^\pm} = 800\ {\rm GeV}$, $\omega_{1,2}=0$, and use the Type II 2HDM.  Ref.~\cite{Haber:2015pua} provides a convenient form for scanning over input parameters for the 2HDM, which we  adopt in this numerical study, we use input parameters $Z_{5,7}$ in place of $m_A$ and $m_{12}^2$ given by
\begin{eqnarray}
m_{12}^2&=&\frac{\sin(2\beta)}{2}\left(m_H^2\sin^2(\beta-\alpha)+m_h^2\cos^2(\beta-\alpha)+\frac{1}{2}\tan(2\beta)(Z_6-Z_7)v^2\right), \nonumber\\
m_A^2 &=& m_H^2\sin^2(\beta-\alpha)+m_h^2\cos^2(\beta-\alpha)-Z_5v^2, \nonumber \\
Z_6 &=&\frac{(m_h^2-m_H^2)\sin(\beta-\alpha)\cos(\beta-\alpha)}{v^2},
\end{eqnarray}
For this set of parameters we obtain the following constraints from unitarity,
\begin{eqnarray}
&&0.42\lsim \tan\beta \lsim 2.4 \nonumber\\
&&-24.5\lsim  \frac{1}{2}\left(17 \mu_3 +13 \mu_4 +13 \mu_6\right)   \lsim 24.5 \nonumber \\
&&-3.8\lsim \kappa_1\lsim 8.0
\end{eqnarray}
In addition the parameters $\nu_1$ and $\nu_2$ as well as  $\nu_1$ and $\kappa_1$ exhibit the correlated unitarity constraint shown in Figure~\ref{example}.
The allowed parameter region for this example in the $\tan\beta-Z_7$ plane is shown in the left panel in Figure~\ref{example}. 
From one-loop Higgs decays at $1\sigma$ we find $|\kappa_1| \lsim 12.4$ as well as the blue dotted areas in Figure~\ref{example}.

To illustrate  the tree-level unitarity constraints implied by Eq.~\ref{unicon} and the constraints from the LHC data fit more generally, we randomly scanned the parameter space of the 2HDM (and its colour-octet extension) to find a set of allowed points.  To produce these figures we have used the custodial symmetry results by Method I as in Eq.~\ref{met1}, including $m_{H^\pm}=m_A$. We have scanned over the range $600\leq M_H\leq 900$~GeV. Our plots reproduce those of Ref.~\cite{Haber:2015pua} for $m_H=300,600$~GeV and we also find that the allowed region is reduced as $m_H$ increases. 
We further scan $Z_{5,7}$ over the ranges $-10\leq Z_5 \leq 2.5$, $-10\leq Z_7\leq 10$. The upper bound on $Z_5$ arises from the requirement of $m_A$ being larger than about 400~GeV \cite{Hermann:2012fc}\footnote{Taking at face value the constraint from $B\to X_s \gamma$: $m_{H^\pm}\geq 380$~GeV.}, and the lower bound keeps $m_A$ below around 1300~GeV. 
$\tan\beta$ is scanned over the range $0.2,50$ and $\cos(\beta-\alpha)$ is scanned over $(-0.5,0.5)$. The charged Higgs mass is equal to $m_A$ and as calculated from Eq.~\ref{masses}, is found to lie in the range  $(400, 1200)$ for these parameter values. The independent parameters that involve the colour octet scalars in the $SU(2)_C$ limit are allowed to vary in the range $-5\pi\leq \nu_{1,2},\omega_{1,2},\kappa_{1,2}\leq 5\pi$, to cover the region implied by Eq.~\ref{approxlim}. The parameters that affect only colour-octet self interactions at tree-level, $\mu_i$ are constrained by Eq.~\ref{mu1con} (which we reproduce numerically by first setting a slightly larger range) and Eq.~\ref{mu2con} which also constrains $\nu_{4,5}, \omega_{4,5}$ which do not affect two-to-two scattering in the colour singlet zeroth partial wave. Finally, the mass $M_{S^\pm}$ is set to 1~TeV, which combined with the other parameters implies  $725 \leq M_{S^0_{R}}\leq 1200$~GeV.

\subsection{Two Higgs doublet model parameters}

\begin{figure}[htb]
\includegraphics[width=1\textwidth]{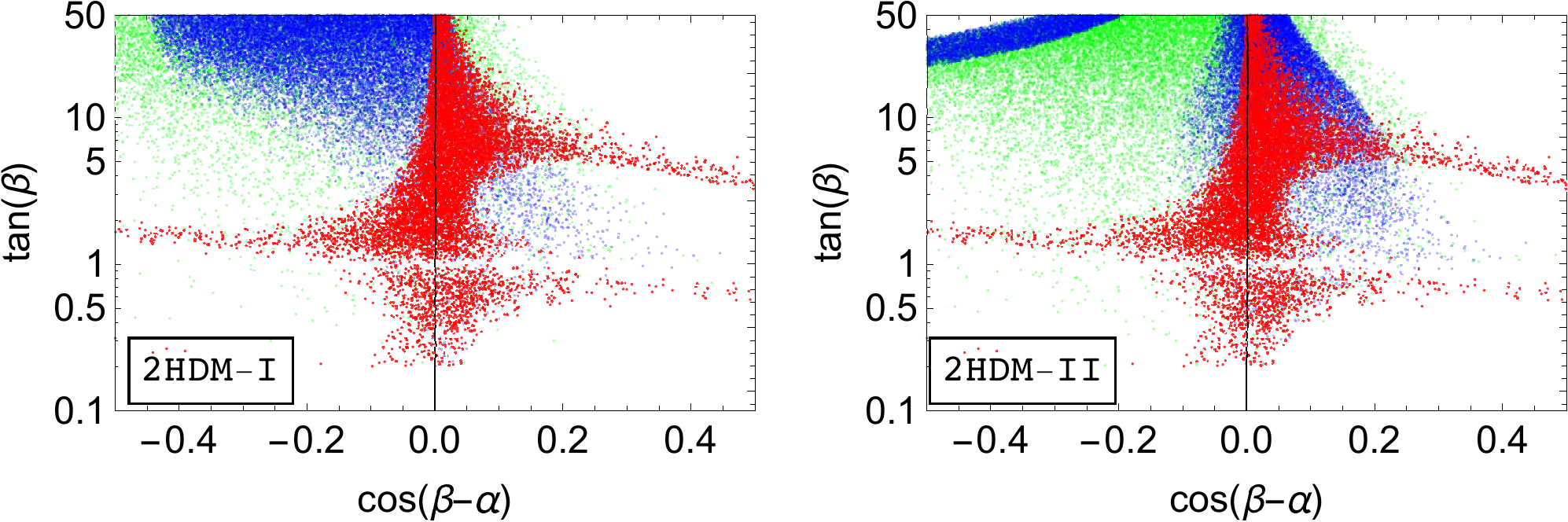}
\caption{
Comparison of unitarity constraints (red points) to  $1\sigma$ constraints  from $h\to gg$ and $h \to \gamma\gamma$ in the 2HDM (blue points) and the 2HDM plus a colour octet (green) as described in the text.}
\label{f:unit} 
\end{figure}

We reproduce the known shape of the region allowed by unitarity  in the $\tan\beta-\cos(\beta-\alpha)$ plane \cite{Haber:2015pua}\footnote{We use the condition $\left|a^0_0\right| \leq \frac{1}{2}$ instead of $|a^0_0|\leq 1$}: it is very narrow for $\tan\beta$ larger than about 10 as can be seen in Figure~\ref{f:unit} and it gets smaller as $M_H$ increases, so that the red region shown is mostly determined by the value $m_H=600$~GeV, the lowest in our range. The same figure shows that there is a small overlap between the regions allowed by unitarity (red) and those allowed by the effective loop decays of the Higgs (blue) in both type-I and type-II 2HDM but this overlap region is enlarged with the addition of the colour octet (green). However, the colour octet tends to populate regions  that are not allowed by the tree-level unitarity constraints. 

\begin{figure}[thb]
\includegraphics[width=1\textwidth]{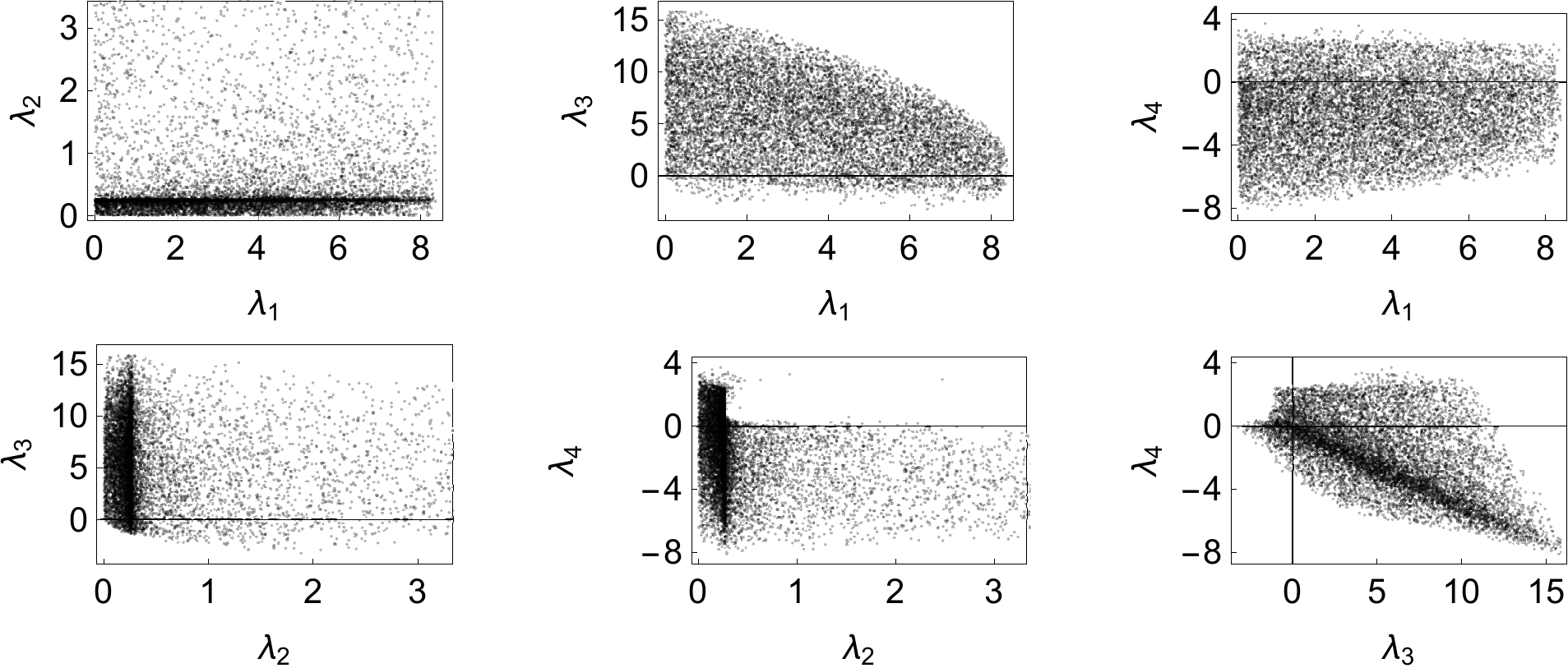}
\caption{
\label{f:unitl} Two dimensional projections of unitarity constraints in 2HDM. }
\end{figure}

\begin{figure}[thb]
\includegraphics[width=1\textwidth]{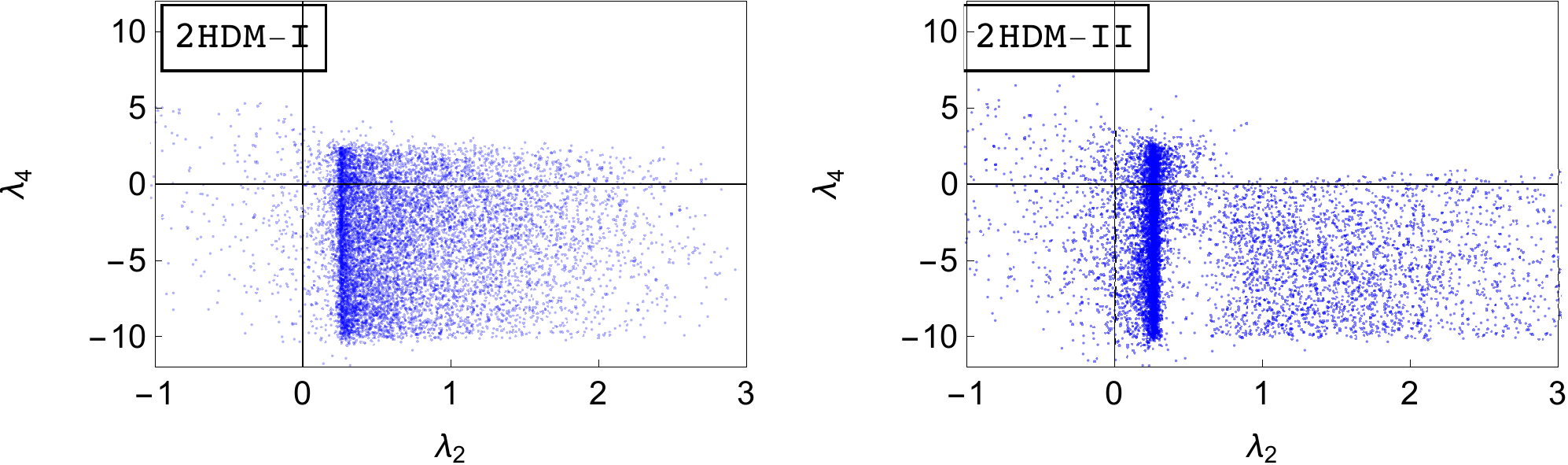}
\caption{
\label{f:loop-uni} Two dimensional projections of the region allowed by $h\to gg$ and $h \to \gamma\gamma$ at one-sigma. }
\end{figure}

Next, we illustrate in Figure~\ref{f:unitl} the two dimensional projections of the multidimensional region allowed by the tree-level unitarity constraints in the parameters of the 2HDM. The more significant correlation found is that between $\lambda_3$ and $\lambda_4$. The darker regions in the plots reflect the concentration of points in the narrow region allowed in the $\tan\beta-\cos(\beta-\alpha)$ plane.
We considered the question of overlap between the allowed regions in Figure~\ref{f:unitl} and additional constraints arising from the one-loop Higgs decays, and found that tree-level unitarity is more restrictive in all cases. We show in Figure~\ref{f:loop-uni} the region most constrained by $h\to gg$ and $h \to \gamma\gamma$.

\subsection{Parameters that mix the 2HDM sector with the colour-octet sector} 

\begin{figure}[thb]
\includegraphics[width=1\textwidth]{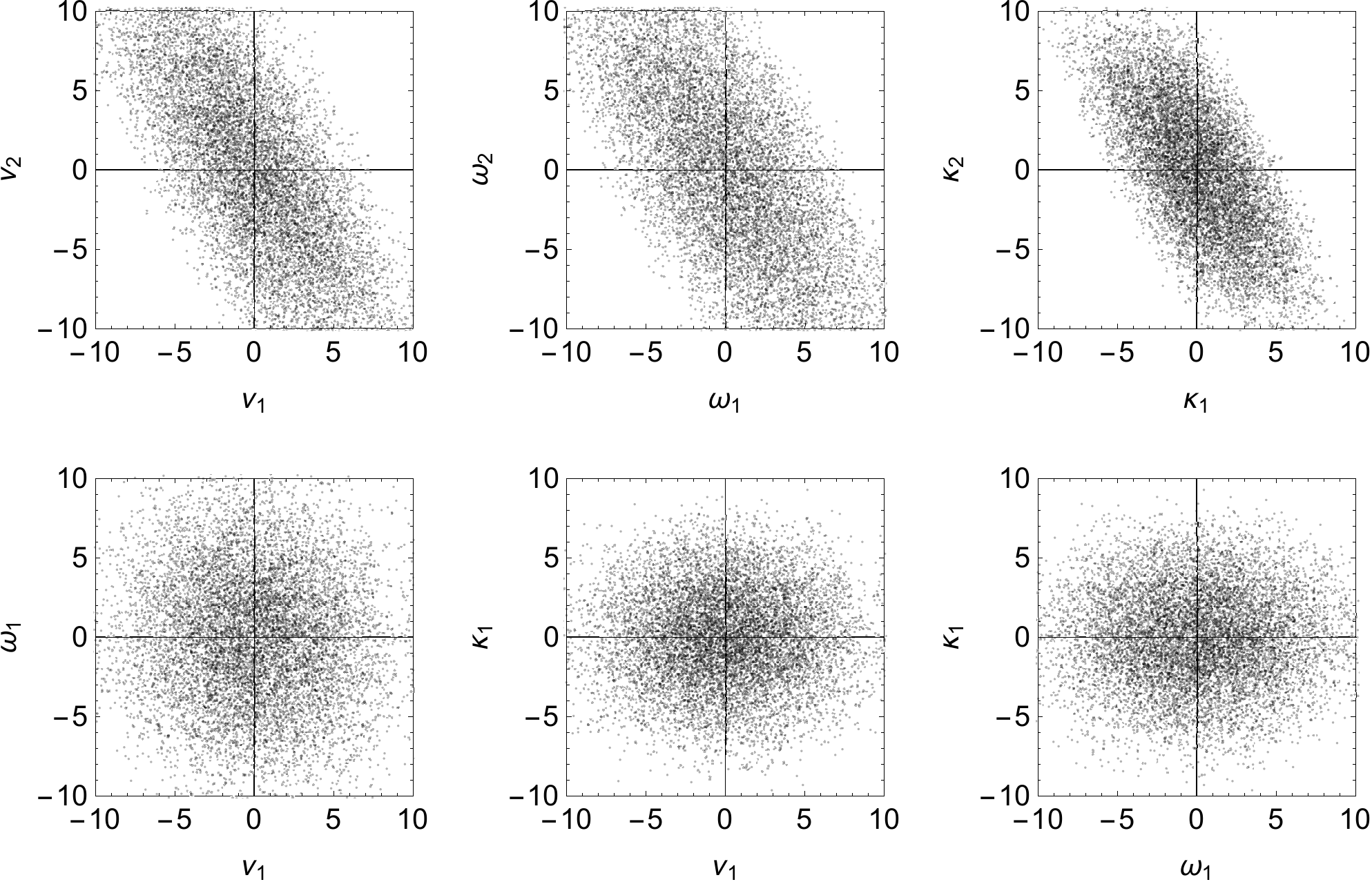}
\caption{
\label{f:unitoctet} Two dimensional projections of unitarity constraints on the parameters that mix the 2HDM scalars and the colour-octet scalars. }
\end{figure}

\begin{figure}[thb]
\includegraphics[width=1\textwidth]{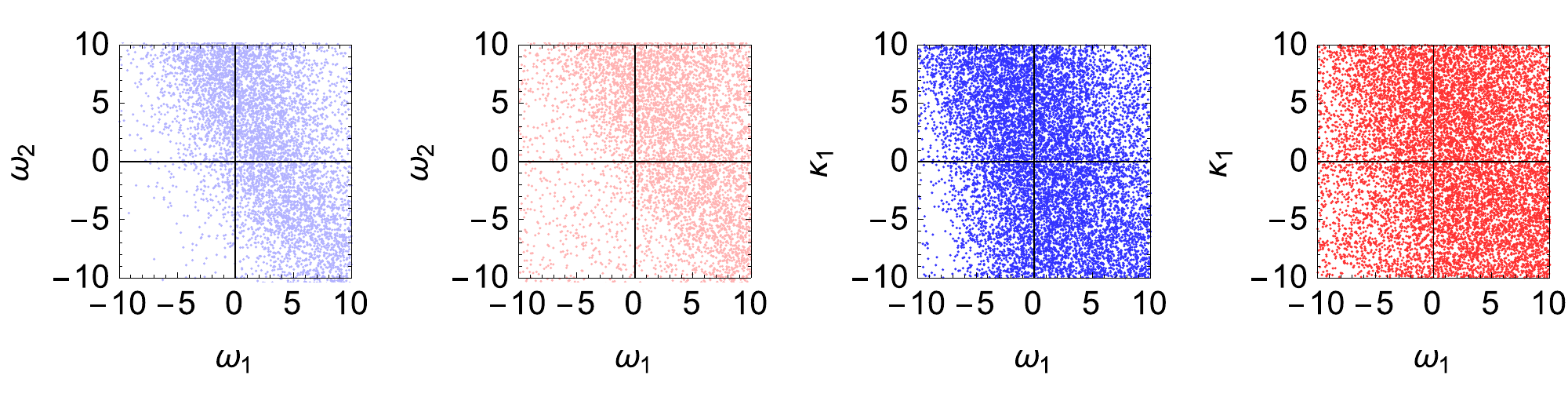}
\caption{
\label{f:loop-uni2} Two dimensional projections of constraints arising from $1\sigma$ allowed regions in $h\to gg$ and $h \to \gamma\gamma$ for 2HDM-I (blue) and 2HDM-II (red). }
\end{figure}

The two dimensional projections of the region allowed by tree-level unitarity for this sector are shown in Figure~\ref{f:unitoctet}. The figures show approximate correlations of the form $|2\nu_1+\nu_2|\lsim 14$, $|2\omega_1+\omega_2|\lsim 15$ and $|2\kappa_1+\kappa_2|\lsim 11$.
In the same manner we study the two-dimensional projections of the region allowed at $1\sigma$ by the loop induced Higgs decays. The only projections indicating a possible correlation are shown in Figure~\ref{f:loop-uni2}. 

\subsection{Loop-induced Higgs decay}

\begin{figure}[thb]
\includegraphics[width=1\textwidth]{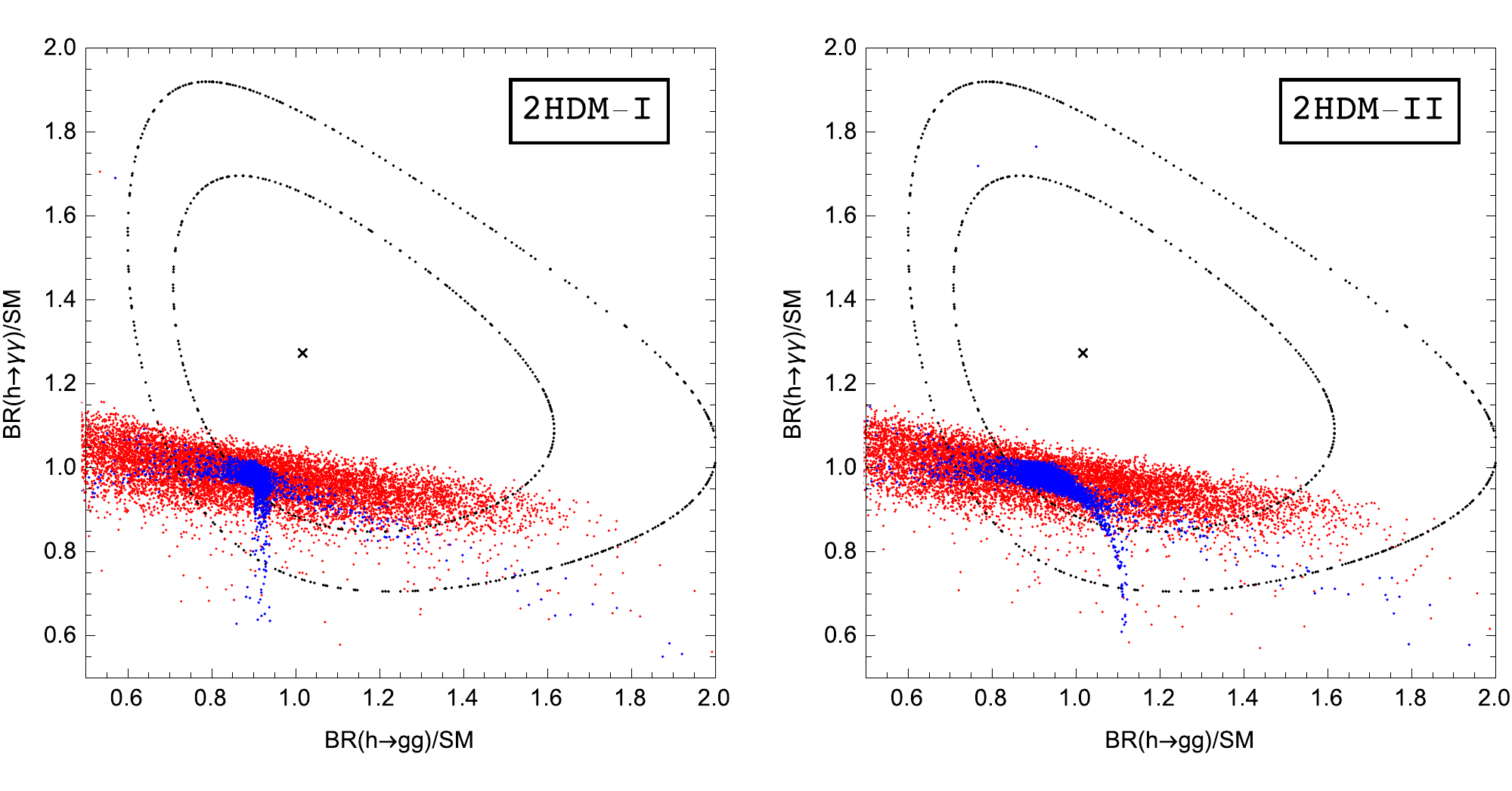}
\caption{
\label{f:hloop} Points in parameter space that satisfy the unitarity constraints shown in a $h\to gg$-$h \to \gamma\gamma$ plot. The blue points correspond to 2HDM whereas the red points correspond to the extended 2HDM. }
\end{figure}

Now we present the points allowed by tree-level unitarity in a $h\to gg$-$h \to \gamma\gamma$ plot in Figure~\ref{f:hloop}. The black contours are taken from Ref.~\cite{Giardino:2013bma}\footnote{We thank Kristjan Kannike who provided us with these fits.} and are respectively the $1\sigma$ and $2\sigma$ allowed regions, with the cross being the best fit point. The SM point is, of course, (1,1). On these contours we have overlaid the blue regions which consist of the points allowed by unitarity for the 2HDM parameter space, and the red regions corresponding to those allowed by unitarity for the 2HDM augmented by the colour-octet.
The colour-octet extends the region which can be explained with a 2HDM mostly in the direction of a larger $BR(h\to gg)$. This figure does not give any insight into the values of different parameters in various regions of the plot. We have studied this issue by looking at all the possible correlations between pairs of parameters and the value of the ($h\to gg$, $h \to \gamma\gamma$) point in Figure~\ref{f:hloop}, but found no notable correlations beyond those already shown in Figure~\ref{f:loop-uni2}. Given the complexity of Eq.~\ref{extraloop} this is not too surprising. One could also constrain the points illustrated in this figure by requiring them to lie within
 the 95\% confidence level region of Figure~\ref{tree-con}. Since this is only an approximation to the global fit, it is easier to  require instead  that they satisfy $-0.04\leq \cos(\beta-\alpha)\leq 0.08$ and $0.1\leq \tan\beta\leq 5$, roughly mapping the region shown in Figure~1 of Ref.~\cite{Dorsch:2016tab} for 2HDM-II. The result is indistinguishable from the red region already in Figure~\ref{f:hloop}. 
These results illustrate how the loop induced Higgs decays are at present the best channels to constrain a Manohar-Wise type colour-octet.

\begin{figure}[thb]
\includegraphics[width=1\textwidth]{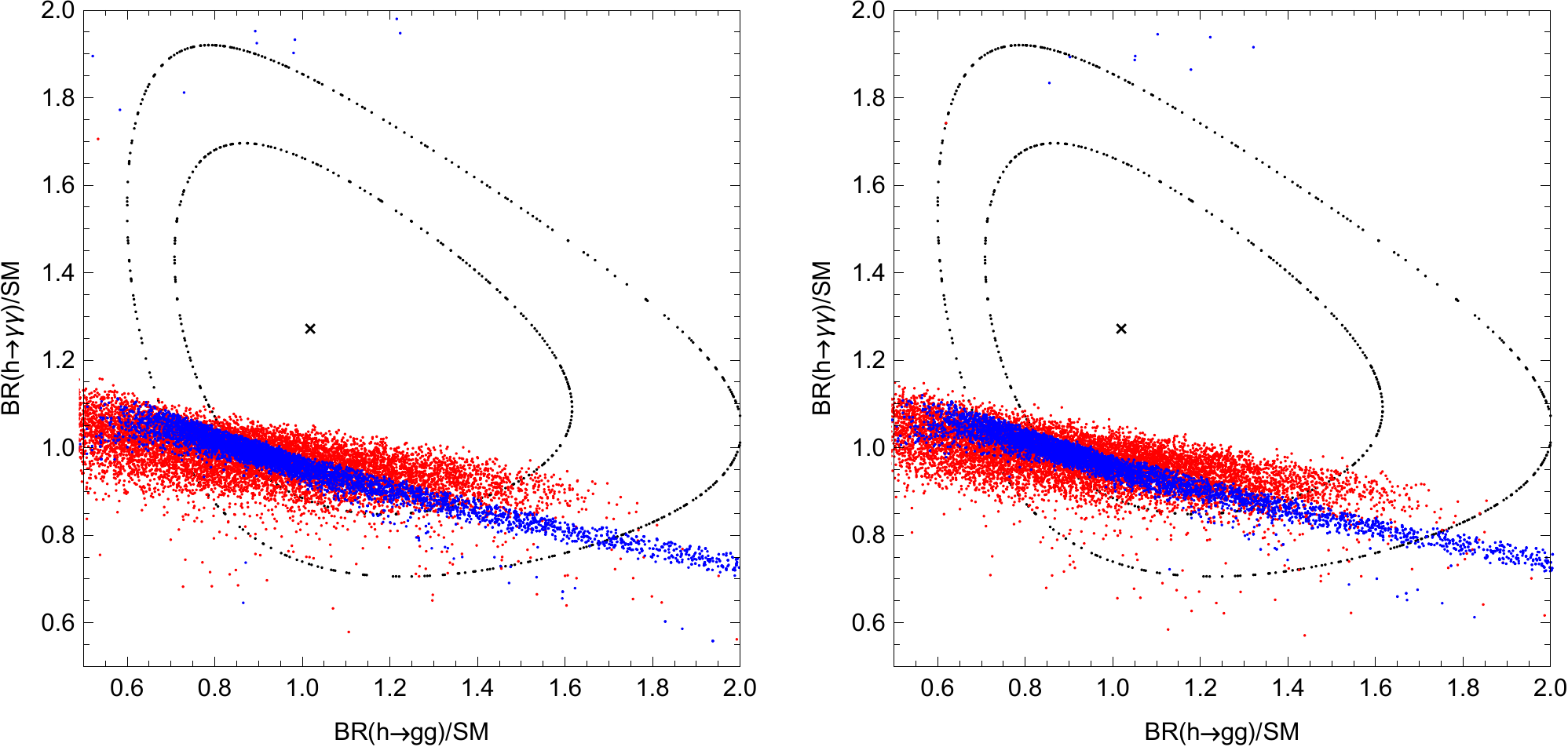}
\caption{
\label{f:hloop2} Points in parameter space that satisfy the unitarity constraints of the extended 2HDM are shown in a $h\to gg$-$h \to \gamma\gamma$ plot. The red points correspond to the $h\to gg, \gamma\gamma$ rates being calculated in the full, colour octet augmented, model. The blue points correspond to the $h\to gg, \gamma\gamma$ rates being calculated without the contributions from the colour octet.}
\end{figure}

We can consider the effect of the additional parameters from the colour-octet sector as follows. For each of the points in parameter space that satisfies the tree-level unitarity constraints we can compute two different points ($h\to gg$, $h \to \gamma\gamma$). The first one would use the results of the 2HDM ignoring the additional contributions from the colour octet. These points are shown in blue in Figure~\ref{f:hloop2}. The second point (in red) is the one corresponding to the calculation in the full model, already shown in Figure~\ref{f:hloop}.

\begin{figure}[thb]
\includegraphics[width=1\textwidth]{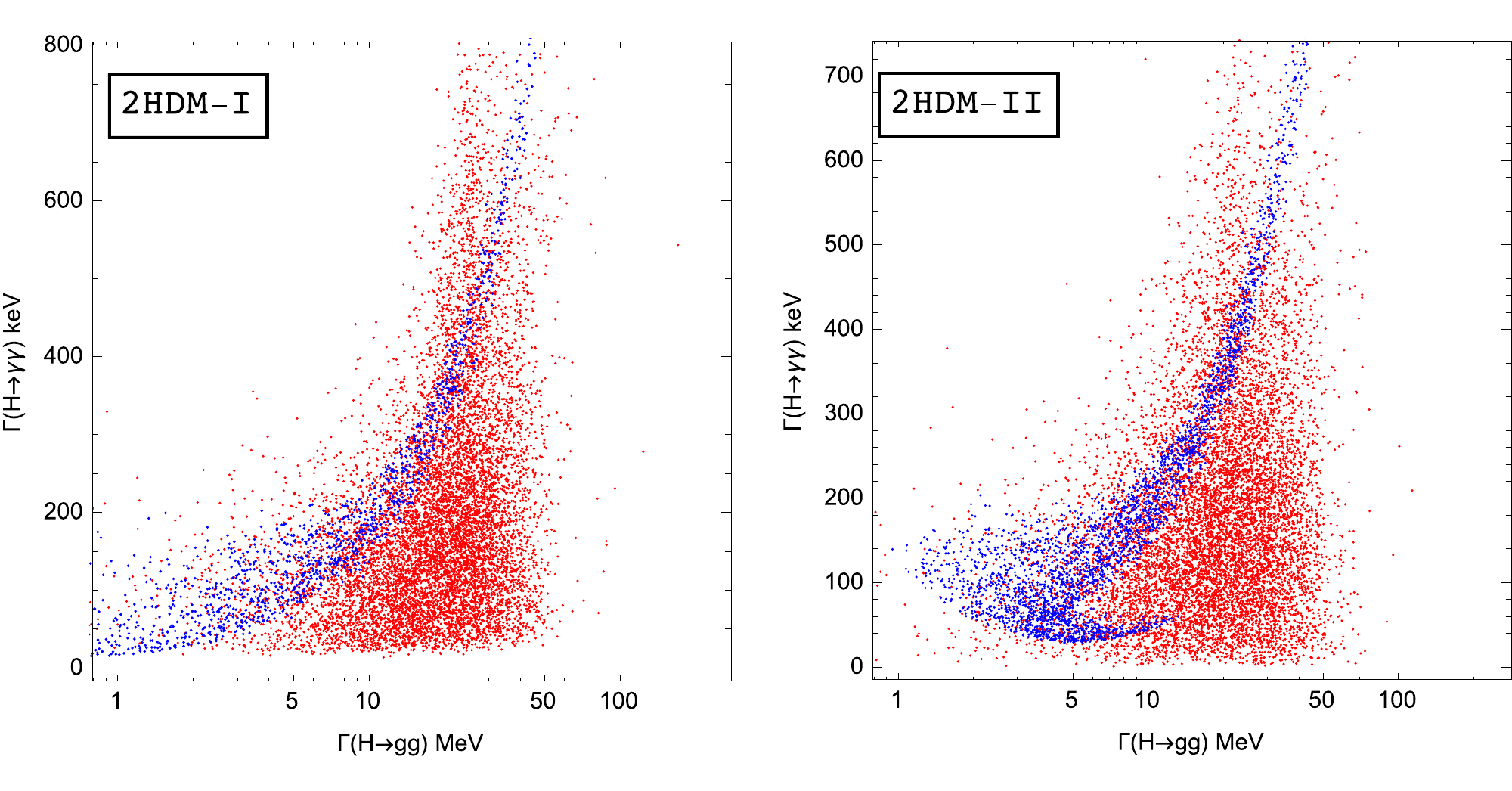}
\caption{
\label{f:Hl} Points in parameter space  with $600\leq M_H \leq 900$~GeV that satisfy the unitarity constraints as well as the $h\to gg$ and $h \to \gamma\gamma$ $1\sigma$ constraints shown in a $H\to gg$-$H \to \gamma\gamma$ plot. The blue points correspond to 2HDM whereas the red points correspond to the extended 2HDM.}
\end{figure}

The region allowed by both tree-level unitarity and Higgs decays at one-loop can be used to predict the loop-induced decays of the heavier neutral scalars. As an example we  show in Figure~\ref{f:Hl} the decay rates for the heavy neutral scalar of the 2HDM, $H^0$, into two photons and two gluons.

\section{Summary and Conclusions}

We have constructed an extension of 2HDM in which a colour-octet electroweak-doublet (MW) is added. Starting from the most general renormalizable scalar potential we have reduced the number of allowed terms with the usual theoretical requirements of minimal flavour violation and custodial symmetry. We have scanned the remaining parameter space to find the region which satisfies perturbative unitarity and have presented two dimensional projections of this region. The high energy two-to-two scattering matrix elements imply that correlations exist between certain pairs of the new couplings which are observed in these projections.

We have then confronted the model with available LHC results in the form of fitted couplings of the Higgs boson which we identify with the lightest scalar in the 2HDM. After collecting constraints on the parameters of the 2HDM from tree-level Higgs couplings we constrain the new sector couplings to the colour-octet using a current fit on the one loop $h\to \gamma\gamma$ and $h\to gg$ couplings.

Addition of the colour-octet affects most the one loop $h\to \gamma\gamma$ and $h\to gg$ modes where it enlarges the allowed region of parameter space in the $\tan\beta-\cos(\beta-\alpha)$ plane, but not notably in the overlap zone with tree-level unitarity constraints as seen in Figure~\ref{f:unit}.  Of course, introducing a new colour-octet scalar doesn't populate more points in the unitarity allowed region when projected to the 2HDM parameter space.

The colour-octet also enlarges the region of overlap with the $1\sigma$ bounds $h\to \gamma\gamma$ and $h\to gg$, 
but the branching ratio of $h \to gg$ tends to increase more significantly than that of $h \to \gamma\gamma$ as can be seen in Figure~\ref{f:hloop}.  

Finally we predict the one loop couplings of the heavier neutral scalar $H\to \gamma\gamma$ and $H\to gg$ using the points in parameter space that satisfy all our constraints.

\begin{acknowledgments}

This research was supported in part by the DOE under contract number DE-SC0009974. Li Cheng thanks Margarida Rebelo for useful correspondence on 2HDM and we thank Kristjan Kannike who provided us with the fits from Ref.~\cite{Giardino:2013bma}.

\end{acknowledgments}


\clearpage

\begin{thebibliography}{999}

\bibitem{Aad:2012tfa} 
  G.~Aad {\it et al.}  [ATLAS Collaboration],
  Phys.\ Lett.\ B {\bf 716}, 1 (2012)
  [arXiv:1207.7214 [hep-ex]].
  
\bibitem{Chatrchyan:2012ufa} 
  S.~Chatrchyan {\it et al.}  [CMS Collaboration],
  Phys.\ Lett.\ B {\bf 716}, 30 (2012)
  [arXiv:1207.7235 [hep-ex]].

\bibitem{atl-cms} 
  The ATLAS and CMS Collaborations,
  ATLAS-CONF-2015-044.

\bibitem{Celis:2013rcs} 
  A.~Celis, V.~Ilisie and A.~Pich,
  JHEP {\bf 1307}, 053 (2013)
  doi:10.1007/JHEP07(2013)053
  [arXiv:1302.4022 [hep-ph]].

\bibitem{Krawczyk:2013gia} 
  M.~Krawczyk, D.~Sokoowska and B.~wieewska,
  J.\ Phys.\ Conf.\ Ser.\  {\bf 447}, 012050 (2013)
  doi:10.1088/1742-6596/447/1/012050
  [arXiv:1303.7102 [hep-ph]].

\bibitem{Barroso:2013zxa} 
  A.~Barroso, P.~M.~Ferreira, R.~Santos, M.~Sher and J.~P.~Silva,
  arXiv:1304.5225 [hep-ph].

\bibitem{Ferreira:2013qua} 
  P.~M.~Ferreira, R.~Santos, M.~Sher and J.~P.~Silva,
  arXiv:1305.4587 [hep-ph].
  

  
\bibitem{Dorsch:2013wja} 
  G.~C.~Dorsch, S.~J.~Huber and J.~M.~No,
  JHEP {\bf 1310}, 029 (2013)
  doi:10.1007/JHEP10(2013)029
  [arXiv:1305.6610 [hep-ph]].

\bibitem{Celis:2013ixa} 
  A.~Celis, V.~Ilisie and A.~Pich,
  JHEP {\bf 1312}, 095 (2013)
  doi:10.1007/JHEP12(2013)095
  [arXiv:1310.7941 [hep-ph]].

\bibitem{Chang:2013ona} 
  S.~Chang, S.~K.~Kang, J.~P.~Lee, K.~Y.~Lee, S.~C.~Park and J.~Song,
  JHEP {\bf 1409}, 101 (2014)
  doi:10.1007/JHEP09(2014)101
  [arXiv:1310.3374 [hep-ph]].

\bibitem{Dumont:2014wha} 
  B.~Dumont, J.~F.~Gunion, Y.~Jiang and S.~Kraml,
  Phys.\ Rev.\ D {\bf 90}, 035021 (2014)
  doi:10.1103/PhysRevD.90.035021
  [arXiv:1405.3584 [hep-ph]].

\bibitem{Ferreira:2014qda} 
  P.~M.~Ferreira, R.~Guedes, J.~F.~Gunion, H.~E.~Haber, M.~O.~P.~Sampaio and R.~Santos,
  arXiv:1410.1926 [hep-ph].


\bibitem{Baglio:2014nea} 
  J.~Baglio, O.~Eberhardt, U.~Nierste and M.~Wiebusch,
  Phys.\ Rev.\ D {\bf 90}, no. 1, 015008 (2014)
  doi:10.1103/PhysRevD.90.015008, [arXiv:1403.1264 [hep-ph]].

\bibitem{Haber:2015pua} 
  H.~E.~Haber and O.~Stal,
  Eur.\ Phys.\ J.\ C {\bf 75}, no. 10, 491 (2015)
  doi:10.1140/epjc/s10052-015-3697-x
  [arXiv:1507.04281 [hep-ph]].

\bibitem{Manohar:2006ga} 
  A.~V.~Manohar and M.~B.~Wise,
  Phys.\ Rev.\ D {\bf 74}, 035009 (2006)
  [hep-ph/0606172].


\bibitem{Gresham:2007ri} 
  M.~I.~Gresham and M.~B.~Wise,
  Phys.\ Rev.\ D {\bf 76}, 075003 (2007)
  [arXiv:0706.0909 [hep-ph]].
  
\bibitem{Gerbush:2007fe} 
  M.~Gerbush, T.~J.~Khoo, D.~J.~Phalen, A.~Pierce and D.~Tucker-Smith,
  Phys.\ Rev.\ D {\bf 77}, 095003 (2008)
  [arXiv:0710.3133 [hep-ph]].



\bibitem{Burgess:2009wm} 
  C.~P.~Burgess, M.~Trott and S.~Zuberi,
  JHEP {\bf 0909}, 082 (2009)
  [arXiv:0907.2696 [hep-ph]].
  
\bibitem{Carpenter:2011yj} 
  L.~M.~Carpenter and S.~Mantry,
  Phys.\ Lett.\ B {\bf 703}, 479 (2011)
  [arXiv:1104.5528 [hep-ph]].
  
\bibitem{Enkhbat:2011qz} 
  T.~Enkhbat, X.~-G.~He, Y.~Mimura and H.~Yokoya,
  JHEP {\bf 1202}, 058 (2012)
  [arXiv:1105.2699 [hep-ph]].

\bibitem{He:2011ti} 
  X.~-G.~He and G.~Valencia,
  Phys.\ Lett.\ B {\bf 707}, 381 (2012)
  [arXiv:1108.0222 [hep-ph]].

\bibitem{Dobrescu:2011aa} 
  B.~A.~Dobrescu, G.~D.~Kribs and A.~Martin,
  Phys.\ Rev.\ D {\bf 85}, 074031 (2012)
  [arXiv:1112.2208 [hep-ph]].
  
\bibitem{Bai:2011aa} 
  Y.~Bai, J.~Fan and J.~L.~Hewett,
  JHEP {\bf 1208}, 014 (2012)
  [arXiv:1112.1964 [hep-ph]].
 
\bibitem{Arnold:2011ra} 
  J.~M.~Arnold and B.~Fornal,
  Phys.\ Rev.\ D {\bf 85}, 055020 (2012)
  [arXiv:1112.0003 [hep-ph]].
 
\bibitem{He:2011ws} 
  X.~-G.~He, G.~Valencia and H.~Yokoya,
  JHEP {\bf 1112}, 030 (2011)
  [arXiv:1110.2588 [hep-ph]].


\bibitem{Cacciapaglia:2012wb} 
  G.~Cacciapaglia, A.~Deandrea, G.~D.~La Rochelle and J.~-B.~Flament,
  arXiv:1210.8120 [hep-ph].
 
\bibitem{Dorsner:2012pp} 
  I.~Dorsner, S.~Fajfer, A.~Greljo and J.~F.~Kamenik,
  arXiv:1208.1266 [hep-ph].
  
\bibitem{Kribs:2012kz} 
  G.~D.~Kribs and A.~Martin,
  arXiv:1207.4496 [hep-ph].
  
    
  
\bibitem{Reece:2012gi} 
  M.~Reece,
  arXiv:1208.1765 [hep-ph].
  
\bibitem{Cao:2013wqa} 
  J.~Cao, P.~Wan, J.~M.~Yang and J.~Zhu,
  arXiv:1303.2426 [hep-ph].

\bibitem{He:2013tla} 
  X.~G.~He, H.~Phoon, Y.~Tang and G.~Valencia,
  JHEP {\bf 1305}, 026 (2013)
  doi:10.1007/JHEP05(2013)026
  [arXiv:1303.4848 [hep-ph]].

  
\bibitem{He:2013tia} 
  X.~G.~He, Y.~Tang and G.~Valencia,
  Phys.\ Rev.\ D {\bf 88}, 033005 (2013)
  doi:10.1103/PhysRevD.88.033005
  [arXiv:1305.5420 [hep-ph]].

\bibitem{Cheng:2015lsa} 
  X.~D.~Cheng, X.~Q.~Li, Y.~D.~Yang and X.~Zhang,
  J.\ Phys.\ G {\bf 42}, no. 12, 125005 (2015)
  doi:10.1088/0954-3899/42/12/125005
  [arXiv:1504.00839 [hep-ph]].

\bibitem{Buttazzo:2014bka} 
  D.~Buttazzo,
  arXiv:1403.6535 [hep-ph].

\bibitem{He:2014xla} 
  X.~G.~He, G.~N.~Li and Y.~J.~Zheng,
  Int.\ J.\ Mod.\ Phys.\ A {\bf 30}, no. 25, 1550156 (2015)
  doi:10.1142/S0217751X15501560
  [arXiv:1501.00012 [hep-ph]].
\bibitem{Yue:2014tya} 
  J.~Yue,
  Phys.\ Lett.\ B {\bf 744}, 131 (2015)
  doi:10.1016/j.physletb.2015.03.044
  [arXiv:1410.2701 [hep-ph]].

\bibitem{Kobakhidze:2014gqa} 
  A.~Kobakhidze, L.~Wu and J.~Yue,
  JHEP {\bf 1410}, 100 (2014)
  doi:10.1007/JHEP10(2014)100
  [arXiv:1406.1961 [hep-ph]].

\bibitem{Cao:2015twy} 
  J.~Cao, C.~Han, L.~Shang, W.~Su, J.~M.~Yang and Y.~Zhang,
  Phys.\ Lett.\ B {\bf 755}, 456 (2016)
  doi:10.1016/j.physletb.2016.02.045
  [arXiv:1512.06728 [hep-ph]].


\bibitem{Bertolini:2013vta} 
  S.~Bertolini, L.~Di Luzio and M.~Malinsky,
  Phys.\ Rev.\ D {\bf 87}, no. 8, 085020 (2013)
  doi:10.1103/PhysRevD.87.085020
  [arXiv:1302.3401 [hep-ph]].

\bibitem{Perez:2016qbo} 
  P.~Fileviez Perez and C.~Murgui,
  arXiv:1604.03377 [hep-ph].

\bibitem{Chivukula:1987py}
  R.~S.~Chivukula and H.~Georgi,
  Phys.\ Lett.\ B {\bf 188} (1987) 99.
  
\bibitem{D'Ambrosio:2002ex}
  G.~D'Ambrosio, G.~F.~Giudice, G.~Isidori and A.~Strumia,
  Nucl.\ Phys.\ B {\bf 645} (2002) 155
  [hep-ph/0207036].

\bibitem{Sikivie:1980hm} 
  P.~Sikivie, L.~Susskind, M.~B.~Voloshin and V.~I.~Zakharov,
  Nucl.\ Phys.\ B {\bf 173}, 189 (1980).
  doi:10.1016/0550-3213(80)90214-X
  
\bibitem{Pomarol:1993mu} 
  A.~Pomarol and R.~Vega,
  Nucl.\ Phys.\ B {\bf 413}, 3 (1994)
  doi:10.1016/0550-3213(94)90611-4
  [hep-ph/9305272].

\bibitem{Grzadkowski:2010dj} 
  B.~Grzadkowski, M.~Maniatis and J.~Wudka,
  JHEP {\bf 1111}, 030 (2011)
  doi:10.1007/JHEP11(2011)030
  [arXiv:1011.5228 [hep-ph]].


\bibitem{Lee:1977eg} 
  B.~W.~Lee, C.~Quigg and H.~B.~Thacker,
  Phys.\ Rev.\ D {\bf 16}, 1519 (1977).
  
\bibitem{Kanemura:1993hm} 
  S.~Kanemura, T.~Kubota and E.~Takasugi,
  Phys.\ Lett.\ B {\bf 313}, 155 (1993)
  [hep-ph/9303263].

\bibitem{Horejsi:2005da} 
  J.~Horejsi and M.~Kladiva
  Eur.\ Phys.\ J.\ C {\bf 46}, 81 (2006)
  doi:10.1140/epjc/s2006-02472-3
  [hep-ph/0510154].

\bibitem{Ginzburg:2005dt} 
  I.~F.~Ginzburg and I.~P.~Ivanov,
  Phys.\ Rev.\ D {\bf 72}, 115010 (2005)
  doi:10.1103/PhysRevD.72.115010
  [hep-ph/0508020].
  
 

  
\bibitem{Grinstein:2015rtl} 
  B.~Grinstein, C.~W.~Murphy and P.~Uttayarat,
  arXiv:1512.04567 [hep-ph].



\bibitem{Holthausen:2011aa}
  M.~Holthausen, K.~S.~Lim and M.~Lindner,
  JHEP {\bf 1202} (2012) 037
  [arXiv:1112.2415 [hep-ph]].
\bibitem{Degrassi:2012ry} 
  G.~Degrassi, S.~Di Vita, J.~Elias-Miro, J.~R.~Espinosa, G.~F.~Giudice, G.~Isidori and A.~Strumia,
  JHEP {\bf 1208}, 098 (2012)
  [arXiv:1205.6497 [hep-ph]];
\bibitem{vsrelated} 
  C.~-S.~Chen and Y.~Tang,
  JHEP {\bf 1204}, 019 (2012)
  [arXiv:1202.5717 [hep-ph]],
    
\bibitem{EliasMiro:2012ay} 
  J.~Elias-Miro, J.~R.~Espinosa, G.~F.~Giudice, H.~M.~Lee and A.~Strumia,
  JHEP {\bf 1206}, 031 (2012)
  [arXiv:1203.0237 [hep-ph]],
  
\bibitem{Lebedev:2012zw} 
  O.~Lebedev,
  Eur.\ Phys.\ J.\ C {\bf 72}, 2058 (2012)
  [arXiv:1203.0156 [hep-ph]],
  
\bibitem{Rodejohann:2012px} 
  W.~Rodejohann and H.~Zhang,
  JHEP {\bf 1206}, 022 (2012)
  [arXiv:1203.3825 [hep-ph]],
  
\bibitem{Cheung:2012nb} 
  C.~Cheung, M.~Papucci and K.~M.~Zurek,
  JHEP {\bf 1207}, 105 (2012)
  [arXiv:1203.5106 [hep-ph]],
  
\bibitem{Kannike:2012pe} 
  K.~Kannike,
  Eur.\ Phys.\ J.\ C {\bf 72}, 2093 (2012)
  [arXiv:1205.3781 [hep-ph]],
  
\bibitem{Iso:2012jn}
  S.~Iso and Y.~Orikasa,
  PTEP {\bf 2013} (2013) 023B08
  [arXiv:1210.2848 [hep-ph]].

\bibitem{Bezrukov:2012sa}
    F.~Bezrukov, M.~Y.~.Kalmykov, B.~A.~Kniehl and M.~Shaposhnikov,
    JHEP {\bf 1210} (2012) 140
    [arXiv:1205.2893 [hep-ph]];
\bibitem{Tang:2013bz}     
  Y.~Tang,
  arXiv:1301.5812 [hep-ph],
\bibitem{Spencer-Smith:2014woa} 
  A.~Spencer-Smith,
  arXiv:1405.1975 [hep-ph].

\bibitem{Mohapatra:2014qva} 
  R.~N.~Mohapatra and Y.~Zhang,
  JHEP {\bf 1406}, 072 (2014)
  doi:10.1007/JHEP06(2014)072
  [arXiv:1401.6701 [hep-ph]].
\bibitem{Buttazzo:2013uya} 
  D.~Buttazzo, G.~Degrassi, P.~P.~Giardino, G.~F.~Giudice, F.~Sala, A.~Salvio and A.~Strumia,
  JHEP {\bf 1312}, 089 (2013)
  doi:10.1007/JHEP12(2013)089
  [arXiv:1307.3536 [hep-ph]].
\bibitem{Kobakhidze:2013pya} 
  A.~Kobakhidze and A.~Spencer-Smith,
  JHEP {\bf 1308}, 036 (2013)
  doi:10.1007/JHEP08(2013)036
  [arXiv:1305.7283 [hep-ph]].
\bibitem{Grinstein:2013npa} 
  B.~Grinstein and P.~Uttayarat,
  JHEP {\bf 1306}, 094 (2013)
  Erratum:[JHEP {\bf 1309}, 110 (2013)]
  doi:10.1007/JHEP09(2013)110, 10.1007/JHEP06(2013)094
  [arXiv:1304.0028 [hep-ph]].

\bibitem{Coleppa:2013dya} 
  B.~Coleppa, F.~Kling and S.~Su,
  JHEP {\bf 1401}, 161 (2014)
  doi:10.1007/JHEP01(2014)161
  [arXiv:1305.0002 [hep-ph]].

\bibitem{Craig:2015jba} 
  N.~Craig, F.~D'Eramo, P.~Draper, S.~Thomas and H.~Zhang,
  JHEP {\bf 1506}, 137 (2015)
  doi:10.1007/JHEP06(2015)137
  [arXiv:1504.04630 [hep-ph]].

\bibitem{Bernon:2015qea} 
  J.~Bernon, J.~F.~Gunion, H.~E.~Haber, Y.~Jiang and S.~Kraml,
  Phys.\ Rev.\ D {\bf 92}, no. 7, 075004 (2015)
  doi:10.1103/PhysRevD.92.075004
  [arXiv:1507.00933 [hep-ph]].

\bibitem{Dorsch:2016tab} 
  G.~C.~Dorsch, S.~J.~Huber, K.~Mimasu and J.~M.~No,
  Phys.\ Rev.\ D {\bf 93}, no. 11, 115033 (2016)
  doi:10.1103/PhysRevD.93.115033
  [arXiv:1601.04545 [hep-ph]].


\bibitem{Gunion:1989we} 
  J.~F.~Gunion, H.~E.~Haber, G.~L.~Kane and S.~Dawson,
  Front.\ Phys.\  {\bf 80}, 1 (2000).
 
\bibitem{Branco:2011iw} 
  G.~C.~Branco, P.~M.~Ferreira, L.~Lavoura, M.~N.~Rebelo, M.~Sher and J.~P.~Silva,
  Phys.\ Rept.\  {\bf 516}, 1 (2012)
  [arXiv:1106.0034 [hep-ph]].
  
  
\bibitem{Gerard:2007kn} 
  J.-M.~Gerard and M.~Herquet,
  Phys.\ Rev.\ Lett.\  {\bf 98}, 251802 (2007)
  doi:10.1103/PhysRevLett.98.251802
  [hep-ph/0703051 [HEP-PH]].
 
\bibitem{Cervero:2012cx} 
  E.~Cervero and J.~M.~Gerard,
  Phys.\ Lett.\ B {\bf 712}, 255 (2012)
  doi:10.1016/j.physletb.2012.05.010
  [arXiv:1202.1973 [hep-ph]].




 
 
 \bibitem{Alekhin:2012py}
   S.~Alekhin, A.~Djouadi and S.~Moch,
   Phys.\ Lett.\ B {\bf 716} (2012) 214
   [arXiv:1207.0980 [hep-ph]].
   
 \bibitem{Masina:2012tz}
   I.~Masina,
   arXiv:1209.0393 [hep-ph].
  
  
 
\bibitem{Chanowitz:1978uj}
  M.~S.~Chanowitz, M.~A.~Furman and I.~Hinchliffe,
  Phys.\ Lett.\ B {\bf 78} (1978) 285.
  
\bibitem{Marciano:1989ns} 
  W.~J.~Marciano, G.~Valencia and S.~Willenbrock,
  Phys.\ Rev.\ D {\bf 40}, 1725 (1989).
  
\bibitem{Cheng:1973nv} 
  T.~P.~Cheng, E.~Eichten and L.~-F.~Li,
  Phys.\ Rev.\ D {\bf 9}, 2259 (1974).

  
\bibitem{Giardino:2013bma} 
  P.~P.~Giardino, K.~Kannike, I.~Masina, M.~Raidal and A.~Strumia,
  JHEP {\bf 1405}, 046 (2014)
  doi:10.1007/JHEP05(2014)046
  [arXiv:1303.3570 [hep-ph]].
 
\bibitem{Deshpande:1977rw} 
  N.~G.~Deshpande and E.~Ma,
  Phys.\ Rev.\ D {\bf 18}, 2574 (1978).
  doi:10.1103/PhysRevD.18.2574
  

\bibitem{Han:2010rf} 
  T.~Han, I.~Lewis and Z.~Liu,
  JHEP {\bf 1012}, 085 (2010)
  doi:10.1007/JHEP12(2010)085
  [arXiv:1010.4309 [hep-ph]].

\bibitem{Khachatryan:2015dcf} 
  V.~Khachatryan {\it et al.} [CMS Collaboration],
  Phys.\ Rev.\ Lett.\  {\bf 116}, no. 7, 071801 (2016)
  doi:10.1103/PhysRevLett.116.071801
  [arXiv:1512.01224 [hep-ex]].
 
\bibitem{Chatrchyan:2012yca} 
  S.~Chatrchyan {\it et al.} [CMS Collaboration],
  Phys.\ Rev.\ D {\bf 87}, no. 7, 072002 (2013)
  doi:10.1103/PhysRevD.87.072002
  [arXiv:1211.3338 [hep-ex]].

   \bibitem{Ellis:1975ap} 
  J.~R.~Ellis, M.~K.~Gaillard and D.~V.~Nanopoulos,
  Nucl.\ Phys.\ B {\bf 106}, 292 (1976).
  

\bibitem{Ioffe:1976sd} 
  B.~L.~Ioffe and V.~A.~Khoze,
  Sov.\ J.\ Part.\ Nucl.\  {\bf 9}, 50 (1978)
  [Fiz.\ Elem.\ Chast.\ Atom.\ Yadra {\bf 9}, 118 (1978)].

\bibitem{Shifman:1979eb} 
  M.~A.~Shifman, A.~I.~Vainshtein, M.~B.~Voloshin and V.~I.~Zakharov,
  Sov.\ J.\ Nucl.\ Phys.\  {\bf 30}, 711 (1979)
  [Yad.\ Fiz.\  {\bf 30}, 1368 (1979)].
  
\bibitem{Djouadi:2005gi} 
  A.~Djouadi,
  Phys.\ Rept.\  {\bf 457}, 1 (2008)
  [hep-ph/0503172].

 
\bibitem{Hermann:2012fc} 
  T.~Hermann, M.~Misiak and M.~Steinhauser,
  JHEP {\bf 1211}, 036 (2012)
  doi:10.1007/JHEP11(2012)036
  [arXiv:1208.2788 [hep-ph]].
  
  
\end{thebibliography}
\end{document}